\documentclass[letterpaper,journal]{IEEEtran}
\usepackage{amsmath,amsfonts}
\usepackage{algorithmic}
\usepackage{algorithm}
\usepackage{array}
\usepackage[caption=false,font=normalsize,labelfont=sf,textfont=sf]{subfig}

\usepackage{array}
\newcolumntype{L}[1]{>{\raggedright\arraybackslash}m{#1}}

\usepackage[hyphens]{url}
\usepackage{hyperref}

\usepackage{textcomp}
\usepackage{stfloats}
\usepackage{url}
\usepackage{verbatim}
\usepackage{graphicx}
\usepackage{cite}
\usepackage{balance}

\IEEEoverridecommandlockouts
\usepackage{cite}
\usepackage{amsmath,amssymb,amsfonts,amsthm}
\usepackage{algorithmic}
\usepackage{graphicx}
\usepackage{textcomp}
\usepackage{algorithm,algorithmic}
\usepackage{xcolor}
\usepackage{mathtools}
\usepackage{bm}
\usepackage{hyperref}
\usepackage{booktabs} 

\usepackage{longtable}
\begin{document}


\title{Electricity Demand and Grid Impacts of AI Data Centers: Challenges and Prospects}

\author{Xin Chen, Xiaoyang Wang, Ana Colacelli, Matt Lee, Le Xie
\thanks{ X. Chen and X. Wang are with the Department of Electrical and Computer Engineering, Texas A\&M University, USA. A. Colacelli is with the Department of Civil and Environmental Engineering, Texas A\&M University. M. Lee is with the Texas A\&M Energy Institute. 
L. Xie is with the John A. Paulson School of Engineering and Applied Sciences, Harvard University. } 
	
\thanks{ 
This work was supported in part by the National Science Foundation Faculty Early Career Development (NSF CAREER) Program under Award No. 2541998, in part by the Consortium on AI and Large Flexible Load (CALL) at Texas A\&M University, in part by the Salata Institute for Climate and Sustainability at Harvard University, in part by the Belfer Center-SEAS Faculty Research Fund on Technology and Geopolitics, and in part by an Ascend Award from Harvard University’s Frontiers of Innovation for Societal Impact Fund. 
(\emph{Corresponding author: Xin Chen}, email: xin\_chen@tamu.edu). 
} 
}

\maketitle

\begin{abstract}

The rapid growth of artificial intelligence (AI) is driving an unprecedented increase in the electricity demand of AI data centers, raising emerging challenges for electric power grids. Understanding the characteristics of AI data center loads and their interactions with the grid is therefore critical for ensuring both reliable power system operation and sustainable AI development. This paper provides a comprehensive review and vision of this evolving landscape. Specifically, this paper (\romannumeral1) presents an overview of AI data center infrastructure and its key components, (\romannumeral2) examines the key characteristics and patterns of electricity demand across the stages of model preparation, training, fine-tuning, and inference, (\romannumeral3) analyzes the critical challenges that AI data center loads pose to power systems across three interrelated timescales, including long-term planning and interconnection, short-term operation and electricity markets, and real-time dynamics and stability, and (\romannumeral4) discusses potential solutions from the perspectives of the grid, AI data centers, and AI end-users to address these challenges. By synthesizing current knowledge and outlining future directions, this review aims to guide research and development in support of the joint advancement of AI data centers and power systems toward reliable, efficient, and sustainable operation.

\end{abstract}

\begin{IEEEkeywords}
AI data centers, electric load demand, grid impact, emerging challenges, potential solutions.
\end{IEEEkeywords}

\section{Introduction}

\IEEEPARstart{I}{n} recent years, the accelerated advancement of generative artificial intelligence (AI), particularly large language models (LLMs) \cite{zhao2023survey} such as GPT, LLaMA, and Gemini, has fueled explosive growth in the AI industry. This surge is driving the rapid expansion of data center infrastructure and imposing unprecedented pressure on the power grid due to the immense electricity demands of ultra-scale AI workloads. 
For instance, training GPT-3 is estimated to have consumed 1.29 GWh of electricity \cite{patterson2021carbon}, whereas the electricity consumption for training the larger-scale GPT-4 rose dramatically to more than an estimated 50 GWh \cite{extnet_gpt4_energy}, equivalent to nearly 0.1\% of New York City’s annual electricity use. 
According to a recent International Energy Agency (IEA) report \cite{iea2024energyai}, global data centers consumed around 415 TWh of electricity in 2024 (about 1.5\% of total global demand), and their consumption is projected to more than double by 2030 to around 945 TWh, with AI identified as the primary driver of this growth.

AI data centers are the computing facilities that are designed and optimized to execute large-scale AI workloads, such as the training and inference of LLMs, computer vision systems, and other compute-intensive AI applications. 
In contrast to traditional data centers, which primarily provide general-purpose information technology (IT) services, AI data centers are architected to deliver the extreme computational performance required for advanced AI workloads. This capability is 
 enabled by dense configurations of high-performance computing (HPC) hardware, such as Graphics Processing Units (GPUs) \cite{choquette2021nvidia} and Tensor Processing Units (TPUs) \cite{jouppi2023tpu}, along with high-efficiency cooling systems \cite{avelar2023ai}. 
Modern hyperscale AI data centers typically operate with a power demand exceeding 100 MW, and some new campuses are planned to scale to the gigawatt (GW) level. 

At present, most data centers obtain electricity mainly through grid interconnection, while islanded configurations remain largely limited to the planning or early deployment stages.
 AI data centers generally connect to medium- or high-voltage distribution grids, whereas hyperscale facilities may connect directly to transmission networks to meet large power requirements~\cite{semianalysis2024electrical}. 
Consequently, the electricity supply mix for AI data centers is primarily shaped by the regional generation portfolio. For example, in the United States, natural gas supplies more than 40\% of data center electricity, followed by renewables, mainly solar and wind generation, at 24\%, nuclear power at approximately 20\%, and coal at about 15\%~\cite{iea2024energyai}. In China, coal accounts for nearly 70\% of data center electricity supply, followed by renewables at nearly 20\%, nuclear power at close to 10\%, and natural gas providing the remaining share~\cite{iea2024energyai}.  These regional differences indicate that the reliability, carbon emissions, and infrastructure impacts of AI data center expansion depend not only on demand magnitude and temporal profiles, but also on local power grid conditions and generation mixes.

As a result, AI data centers are emerging as prominent large electric loads in the power grid, with distinct characteristics and demand patterns, including:    
\begin{itemize}
    \item \emph{High power density:} Intensive AI workloads require substantially more power than conventional IT services. For instance, a ChatGPT query is estimated to consume about 2.9 Wh, nearly ten times the 0.3 Wh of a regular Google search \cite{aljbour2024powering}. In contrast to traditional server racks typically operating at 7-10 kW, AI computing racks can reach power densities of 30-100+ kW per rack \cite{nlyte_rack_power}. 
Such extreme power density places considerable strain on data center architecture, cooling systems, and power grids.

\item \emph{Fast and large variability:} 
Across the training, fine-tuning, and inference phases, AI workloads can be highly variable and bursty \cite{Li2024Sep}, with power demand fluctuating sharply over very short timescales and remaining difficult to forecast \cite{mughees2025forecast}. For example, large-scale GPU clusters can produce power fluctuations of hundreds of megawatts within only seconds \cite{choukse2025power}, bringing significant challenges for system balancing and reliable grid operation.

\item \emph{Grid interface via power electronics:}
AI data centers interface with the power grid primarily through power electronic converters, which exhibit fundamentally different dynamic behaviors from conventional electromechanical loads, such as low inertia, fast response dynamics, and harmonic distortions \cite{Li2025AILoadDynamics}.  At scale, power electronics-based AI computing loads can pose major threats to grid stability and cause severe power quality issues.

 \item \emph{Geographic concentration:}
 AI data centers tend to cluster in regions with low electricity prices, ample land and water
resources, and favorable policy environments. In the U.S., for example, fifteen states (most notably Virginia, Texas, and California)
accounted for about 80\% of total data center demand in 2023 \cite{aljbour2024powering}.
Such concentration effects further intensify local grid stress, and many regional grids cannot accommodate large-scale integration of data centers without substantial infrastructure upgrades.

\end{itemize}

\begin{figure*}[htbp]
    \centering
      \includegraphics[width=0.9\textwidth]{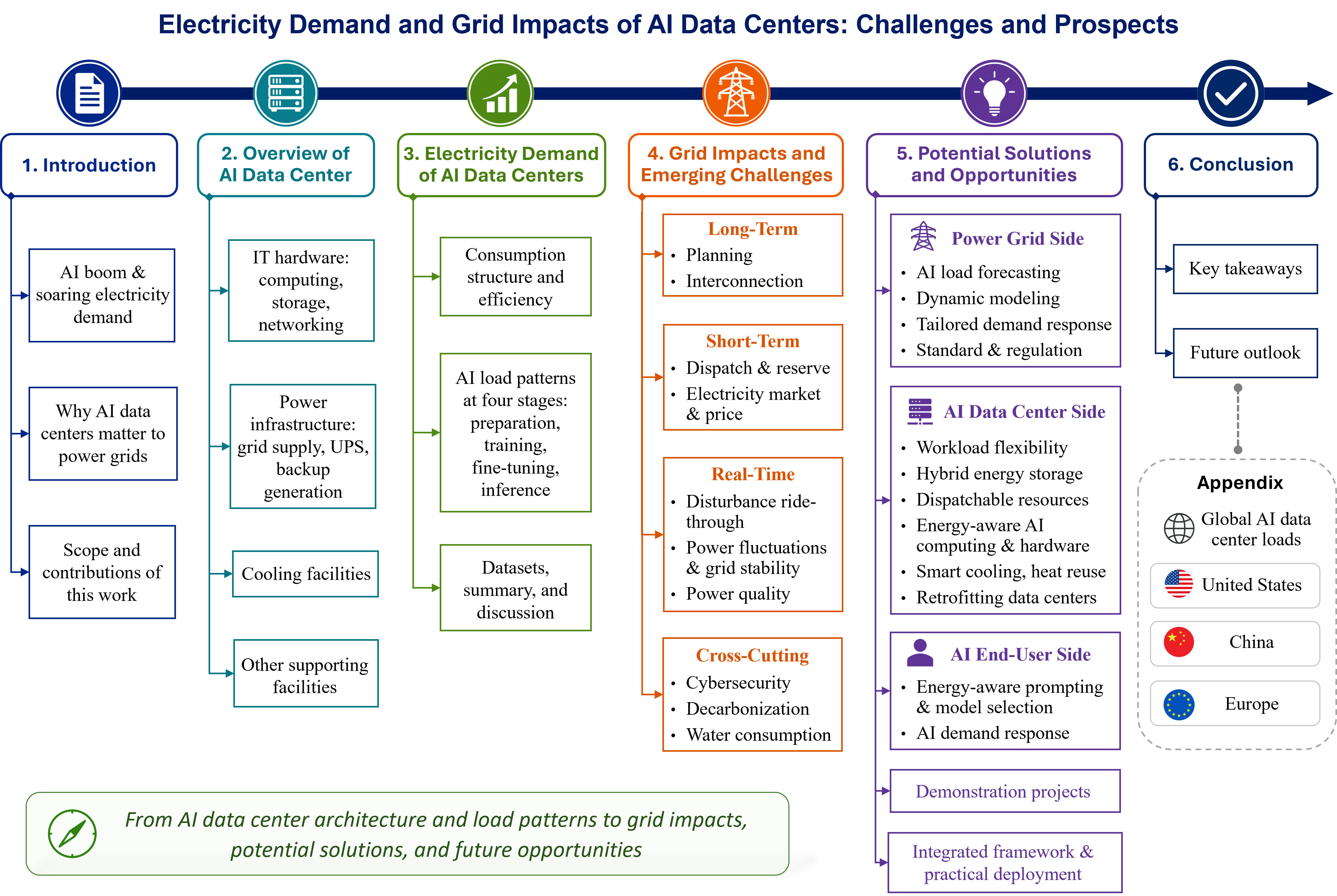}
    \caption{The structure and roadmap of this work.}
    \label{fig:roadmap}
\end{figure*}

As such, the rapid expansion of AI data centers poses unprecedented challenges to modern power systems. To sustain the fast pace of AI innovation and the continued growth of AI infrastructure, it is essential to establish a deep understanding of AI data center loads and their interactions with the grid, and to develop effective solution strategies that ensure reliable and sustainable power system operation. 
A growing number of recent articles and reports \cite{Pilz2025Jan,iea2024energyai,mckinsey2024datacenters,aljbour2024powering,davenport2024generational} have drawn considerable attention to the surging electricity demand of AI data centers and their implications for the power sector.
Prior works \cite{avelar2023ai,aljbour2024powering} 
examine the energy use of AI data centers by analyzing the power demand of key components, including AI computing, cooling systems, and auxiliary facilities, and present  strategies for energy-efficient operation. A
recent study \cite{Li2024Sep} analyzes the transient dynamic behaviors of AI computing power consumption, develops high-level mathematical models for AI data center loads, and discusses the resulting disruptions and opportunities for power systems. Case studies on several power grids are conducted in \cite{lin2024exploding} to assess their capacity to accommodate projected five-year AI data center load growth, deriving implications for long-term grid planning and  management. However, the existing literature has mainly focused on the AI data center itself and presented high-level, preliminary, and fragmented discussions, while a comprehensive and in-depth synthesis of AI data center load characteristics and their impacts on power grids, particularly the challenges from the grid's perspective, remains lacking.

To bridge this critical gap and help guide future research and development in the field, this paper provides a comprehensive review and vision
that (\romannumeral1) details the overall architecture of AI data centers together with their load characteristics and patterns,  (\romannumeral2) systematically analyzes the multi-timescale and multi-faceted challenges posed by the rapid growth of AI data centers to power grids, and (\romannumeral3) presents promising solutions
for enabling their large-scale grid integration in a reliable, sustainable, and grid-supportive manner.

The structure and roadmap of this paper are illustrated in Figure \ref{fig:roadmap}. 
Specifically, Section \ref{sec:overview} presents an overview of AI data center infrastructure and its key components. Section \ref{sec:load} examines the characteristics and patterns of electricity demand across the AI model preparation, training, fine-tuning, and inference stages. Section \ref{sec:gridimpact} analyzes the key challenges that AI data center loads bring to power systems across three interrelated timescales: long-term planning and interconnection, short-term operation and electricity markets, and real-time grid dynamics and stability. Other critical challenges, including cybersecurity,
decarbonization, and water consumption, are discussed as well. Section \ref{sec:solution} proposes potential solutions from the grid side, AI data center side, and AI end-user side to address these challenges and ensure reliable power system operation. Finally, conclusions are drawn in Section \ref{sec:conclusion}.

\section{Overview of an AI Data Center} 
\label{sec:overview}

\begin{figure*}[htbp]
    \centering
    \includegraphics[width=1\textwidth]{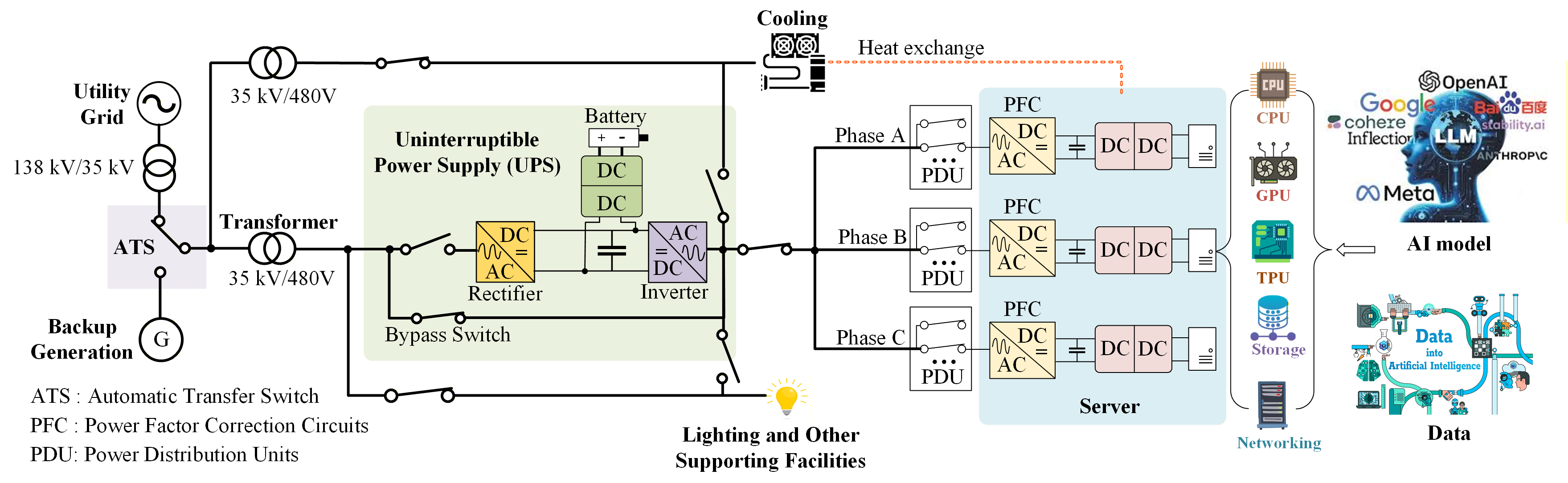}
    \caption{A typical architecture of an AI data center. (The transformer voltage levels and the presented double-conversion UPS system are illustrative examples; actual configurations vary across data centers). }
    \label{fig:structure}
\end{figure*}

This section provides an overview of the primary components of an AI data center, including IT hardware, electrical power infrastructure, cooling systems, and other supporting facilities. A typical architecture illustrating the interconnection of these components is depicted in Figure~\ref{fig:structure}. 

Key distinctions between AI and traditional data centers are summarized in Table~\ref{tab:dc_comparison}. Traditional enterprise data centers typically have relatively low power capacities, often ranging from a few megawatt (MW) to around 10 MW, while large co-location or cloud data centers may reach tens of MW or around 100 MW. Their workloads are generally diversified and smooth, supporting services such as web hosting, enterprise computing, databases, cloud storage, and virtualization. 
 In contrast, AI data centers can reach several hundred MW, with some hyperscale campuses planned at the GW scale. They are designed for large-scale GPU-intensive training, fine-tuning, and inference workloads, which lead to much higher rack power densities, more synchronized workload behavior, sharper ramping events, and burstier power fluctuations. Hence, AI data centers differ from traditional data centers not only in total power capacity, but also in workload characteristics, temporal variability, and dynamic grid-interface behavior, resulting in distinct and more significant grid impacts.

\begin{table*}[htbp]
\centering
\caption{Comparison between traditional data centers and AI data centers}
\label{tab:dc_comparison}
\begin{tabular}{L{2.7cm}L{7cm}L{7cm}}
\toprule
 \textbf{Feature} & \textbf{Traditional Data Center} & \textbf{AI Data Center} \\
\midrule
Primary Functions & General-purpose IT services (e.g., web/app hosting, databases, 
enterprise software hosting, cloud storage, email, virtualization, backup recovery) & AI/ML model training, fine-tuning, and inference (e.g., large language models, AI computer vision, generative AI) \\
\midrule
Workload Pattern & Stable, predictable workloads & Dynamic, bursty, data-intensive, hard-to-predict workloads \\
\midrule
Compute Hardware & CPU-centric, some GPUs & GPU/TPU-dense clusters \\
\midrule
Rack Power Density & 7 kW - 10~kW/rack, moderate density & 30 kW - over 100~kW/rack, very high density \\
\midrule
Storage & Balanced performance and capacity, mix of HDD and SSD & Very high-throughput SSD/NVMe storage with parallel file systems \\
\midrule
Networking & Standard Ethernet, designed for general-purpose traffic patterns & Ultra-high-bandwidth, low-latency interconnects (InfiniBand, NVLink) for distributed AI workloads \\
\midrule
Cooling & Primarily air cooling & Liquid cooling (direct-to-chip, immersion) or hybrid cooling \\
\midrule

Facility Design & Optimized for mixed workloads, standard floor loading & Optimized for high-density AI workloads, reinforced structures for heavy racks and cooling equipment \\
\bottomrule
\end{tabular}
\end{table*}

\subsection{IT Hardware Infrastructure}

The IT hardware infrastructure forms the computational backbone of an AI data center and encompasses the \emph{computing}, \emph{storage}, and \emph{networking} systems, which execute large-scale model training and inference, manage massive datasets, and enable high-throughput communication.

\subsubsection{Computing} AI workloads, such as LLM training and inference, require high-performance computing clusters designed to execute large-scale computations efficiently.
These clusters typically integrate multi-core Central Processing Units (CPUs) with dense configurations of GPUs and TPUs \cite{jouppi2023tpu}. 
These processing units are housed in high-density server nodes and stacked in standardized computer racks, allowing scalable deployment, efficient power distribution, and optimized cooling.
Compared with conventional data centers, AI-oriented computing facilities concentrate substantially greater processing power within each rack. While a standard rack typically uses 7-10 kilowatts (kW), an AI-capable rack can demand 30 kW to over 100 kW \cite{aljbour2024powering, davenport2024generational}, 
with an average of more than 60 kW in dedicated AI facilities \cite{nlyte_rack_power}. For example, the per-rack power of NVIDIA's GB200 NVL36 and NVL72 servers reaches about 66 kW and 120 kW \cite{semianalysisrack}, respectively. 
The substantially higher power densities pose critical challenges for power distribution and thermal management.

\subsubsection{Storage} To sustain intense compute capabilities, scalable and high‑performance data storage is required to handle massive training datasets, intermediate model checkpoints, and inference data. Tiered storage components include: (\romannumeral1) a \emph{hot tier}, such as non-volatile memory express (NVMe) solid-state drive (SSD) arrays, which provides ultra-low latency and high bandwidth for the fast processing of active data; (\romannumeral2) a \emph{warm tier}, such as high-capacity hard disk drives (HDDs), which provides
cost-effective storage for large and less frequently accessed datasets; and (\romannumeral3) a \emph{cold tier}, such as tape libraries or object-storage systems, which
supports long-term archival storage with slower access speeds. 
 In addition, \emph{distributed file systems}, such as Lustre and Ceph, allow large‑scale parallel reading and writing, while \emph{object storage} is used for archiving large datasets with long durability and redundancy \cite{AmazonS32024}. 
Unlike conventional data centers, which balance capacity, cost, and speed, AI-optimized storage prioritizes maximum throughput and parallel access to prevent bottlenecks that idle expensive GPU clusters.

\subsubsection{Networking}
High‑speed networking is critical to ensure that  massive volumes of data flow efficiently between storage systems and compute nodes during large‑scale AI workloads.
The networking system in an AI data center includes both \emph{intra-node interconnects}, such as NVIDIA NVLink~\cite{nvidia_nvlink} for high-bandwidth GPU-to-GPU communication within a server, and \emph{inter-node fabrics}, such as InfiniBand or Ethernet with RDMA over Converged Ethernet (RoCE), for low-latency, high-throughput communication between servers and racks. Specialized network topologies, including fat‑tree and dragonfly architectures \cite{wilke2020opportunities}, are adopted to optimize GPU‑to‑GPU communication in distributed training environments.
In contrast, conventional data centers are usually designed for more heterogeneous workloads and general-purpose traffic patterns, and do not require such high bandwidth or low latency.

\subsection{Electric Power Infrastructure} \label{sec:powerinf}

\subsubsection{Grid Power Supply}

AI data centers are typically connected to the medium- or high-voltage distribution grid, while hyperscale facilities may connect directly to the transmission grid as their primary power source. Depending on the local grid infrastructure and facility scale,  interconnection voltages can range from about 13.2 kV for small-scale installations (under 10 MW) to 115-230 kV for hyperscale sites (exceeding 100 MW) in the U.S.~\cite{semianalysis2024electrical}. 
As illustrated in Figure \ref{fig:structure}, alternating current (AC)
power enters the facility through high‑voltage switchgear and protective devices, after which it is stepped down to operational voltage levels such as 480 V via on‑site three‑phase step‑down transformers. These transformers not only reduce voltage but also provide electrical isolation between the utility grid and the facility’s internal power supply system. 
The power is then distributed through main distribution panels and power distribution units (PDUs) to supply IT server racks, cooling systems, and other supporting loads. For IT server racks, AC power is further converted to direct current (DC) through a power factor correction (PFC) rectifier, followed by a DC-DC converter
for DC-side voltage regulation \cite{sun2022dynamic},
ensuring stable and properly rated power delivery to IT equipment. Additionally,
large AI data centers typically employ redundant utility feeders and transformers to ensure high reliability and continued operation even during partial grid outages \cite{chalise2015data}.

\subsubsection{Uninterruptible Power Supply (UPS)}
UPS systems \cite{muhammad2016review} serve as the first layer of protection against power interruptions in data center operations. They provide instantaneous electrical power from energy storage devices, such as batteries or flywheels, when the grid supply is disrupted, ensuring continuous operation during the transition period before backup generators start and synchronize with the load. UPS deployments commonly employ battery technologies such as valve-regulated lead-acid (VRLA), lithium-ion, and emerging solid-state chemistries \cite{teague2025comparative}. Selection criteria include specific power and energy ratings, cycle life, energy density, thermal management requirements, and cost-effectiveness. 

\begin{figure*}[htbp]
    \centering
    \includegraphics[width=0.9\linewidth]{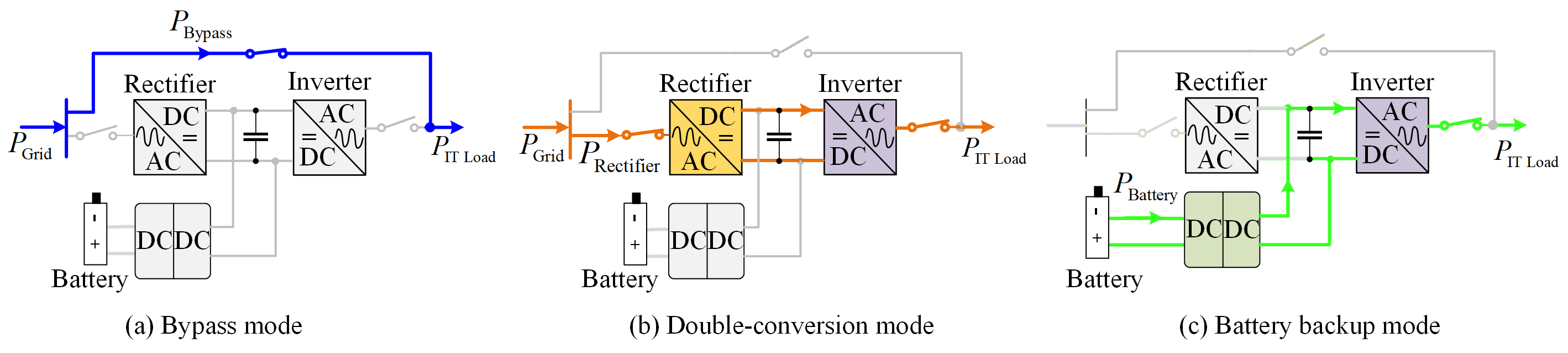}
    \caption{Three operating modes of a typical AI data center with UPS systems.}
    \label{fig:threemode}
\end{figure*}

With UPS systems, a data center typically operates in and switches among three modes: \emph{bypass mode}, \emph{double-conversion mode}, and \emph{battery backup mode} \cite{sun2022dynamic}, as illustrated in Figure~\ref{fig:threemode}. These operating modes are switched according to grid voltage and power conditions to ensure a highly reliable power supply for IT loads. 
One representative mode-transition strategy
is that, when the grid voltage remains within an acceptable range (e.g., 0.9-1.0~pu), a data center operates in bypass mode, with the bypass switch closed and the IT load supplied directly by the grid.
When the grid voltage deteriorates to a moderate range (e.g., 0.7-0.9~pu), the UPS may open the bypass path and operate in double-conversion mode, where grid-side AC power is converted to DC through a rectifier and then back to AC through an inverter with a well-regulated voltage for IT load supply. Under more severe voltage sags (e.g., below 0.7~pu), the UPS may switch to battery backup mode, in which the IT load is supplied by the battery through the inverter until the grid voltage recovers or backup generation becomes available. In Section \ref{sebsec:Real-Time Dynamics and Grid Stability}, Figure~\ref{fig:modeswitch} further illustrates
the simulated operating mode transitions and reconnection behavior of an AI data center during voltage sag and recovery.
Modern UPS systems may also support grid-interactive capabilities, enabling data centers to shift load or participate in demand response programs by temporarily charging and discharging  their stored energy.

\subsubsection{Backup Generation}

Backup generation provides emergency power to ensure that AI workloads remain uninterrupted in the event of a grid outage. Traditionally, diesel generators have been the most common backup solution because of their mature technology, low capital cost, high power density, and fast start-up capability to reliably support the full critical IT load and essential cooling systems. To improve sustainability and reduce local air pollutants and carbon emissions, cleaner backup generation technologies, such as natural gas generators, hydrogen fuel cells, on-site renewable generation, and battery energy storage systems (BESS), are being explored to supplement or partially replace diesel generators~\cite{kambhampati2024moving}. 
Natural gas generators are increasingly considered for AI data center backup because they can provide dispatchable, firm power with lower local pollutant and carbon emissions than diesel generators. They can also be more cost-effective for longer-duration operation when pipeline gas is available. However, since natural gas generators still rely on fossil fuels and may face fuel-delivery constraints during extreme weather or gas system disruptions, they are often viewed as a transitional option or as one component of hybrid backup systems.

Hydrogen fuel cells are emerging as a promising low-emission backup power option for data centers, with recent demonstrations validating their feasibility at the megawatt scale. For example, Microsoft tested a prototype 3 MW hydrogen fuel cell system as an emission-free backup power source for data centers~\cite{microsoft2022hydrogen}. In addition, Caterpillar demonstrated a 1.5 MW hydrogen fuel cell system integrated with battery energy storage at Microsoft’s data center, simulating a 48-hour backup power event while supporting critical data center load~\cite{gooding2024microsoft}. Nevertheless,  hydrogen fuel cell systems currently face higher upfront costs and additional infrastructure requirements, including hydrogen production or delivery, on-site storage, safety systems, and refueling logistics, and thus remain an emerging alternative in data center backup applications. 
On-site solar photovoltaic (PV) generation is another backup option for improving sustainability and reducing carbon emissions. However, PV generation is intermittent and cannot independently provide firm backup power during grid outages, prolonged cloudy periods, or nighttime operation. Therefore, PV systems need to be coordinated with BESS or other dispatchable resources to meet the high reliability requirements of data centers. In such hybrid architectures, BESS can provide fast response, short-duration ride-through, and smoothing of PV fluctuations, while fuel cells or conventional generators can provide longer-duration backup power. Most backup generation systems are equipped with automatic transfer switches or microgrid controllers, enabling multiple resources to be coordinated and rapidly connected to maintain a continuous power supply during outages.

\begin{table*}
\centering
\caption{Summary of Electric Power Infrastructure in AI Data Centers}
\begin{tabular}{@{}L{3.4cm}L{2.5cm}L{4.8cm}L{5.7cm}@{}}
\toprule
\textbf{Infrastructure Component} & \textbf{Function} & \textbf{Key Equipment}& \textbf{Details and Features} \\
\midrule
Grid Power Supply & Primary electricity source & Switch and protective devices, step-down transformers, power distribution units, power converters  & Connected to 13.2-230 kV grid, stepped down to 480/415 V using isolation transformers, redundancy ensures reliability \\
\midrule
Uninterruptible Power Supply (UPS) & Instantaneous power during interruptions & Valve-regulated lead-acid (VRLA), lithium-ion, solid-state batteries, flywheels & Instantaneous power supply until backup starts, protect GPU/TPU racks, support grid-interactive capabilities and demand response \\
\midrule
Backup Generation & Emergency power during grid outages & Diesel generators, solar PV, hydrogen fuel cells, natural gas generators & Support critical IT and cooling loads, automatic transfer switches enable seamless transition, renewable integration \\
\bottomrule
\end{tabular}
\label{tab:power_infrastructure_summary}
\end{table*}

\subsection{Cooling Facilities}

In AI data centers, cooling facilities are essential for dissipating the substantial heat generated by high-density computing equipment, particularly GPU- and TPU-based clusters, maintaining safe operation and preventing hardware failures.
A variety of cooling technologies have been used in data centers \cite{Zhang2021CoolingSurvey}, which are broadly classified into three main categories:
(\romannumeral1) \emph{Air cooling} \cite{ni2017review}
circulates chilled air to absorb heat from IT equipment and expel it from the facility. Due to its low cost and operational simplicity, air cooling
 has been the mainstream choice for data centers with low rack power densities (below 20 kW), but it becomes inadequate for AI clusters with much higher rack power densities~\cite{avelar2023ai}. (\romannumeral2) \emph{Liquid cooling} uses a liquid medium to remove heat from servers, which absorbs heat far more effectively than air and allows for very high rack power densities.
 For example, \emph{liquid immersion cooling} \cite{kuncoro2019immersion} submerges servers in engineered dielectric fluids, enabling safe, direct contact with electronic components and highly efficient heat removal; 
\emph{direct‑to‑chip liquid cooling} \cite{heydari2024experimental} circulates coolant through cold plates mounted directly on processors or other high-heat components, improving heat transfer without immersing the entire server in liquid.  (\romannumeral3) \emph{Hybrid cooling} \cite{chen2019thermodynamic} combines air and liquid cooling methods to optimize efficiency under varying workloads. 
This approach offers flexible operation, reduces energy consumption, and mitigates the high infrastructure cost and complexity of fully liquid-cooled designs.
The commonly used cooling methods are summarized in Table \ref{tab:cooling_comparison}. In practice, cooling selection depends on factors such as rack power density, energy efficiency targets, facility design constraints, capital and operational costs, and water availability. In particular, rapidly growing AI workloads are accelerating the adoption of advanced liquid-based and intelligent cooling solutions. Further discussion on smart and efficient cooling techniques is provided in Section~\ref{sec:centersolu}.

\begin{table*}
\centering
\caption{Comparison of AI Data Center Cooling Techniques}
\begin{tabular}{@{}L{3cm}L{3.3cm}L{1.9cm}L{4cm}L{4cm}@{}}
\toprule
\textbf{Cooling Method} & \textbf{Heat Density Suitability} & \textbf{Efficiency (PUE) } & \textbf{Pros} & \textbf{Cons} \\
\midrule
Air Cooling \cite{ni2017review} & Low-moderate & $\sim$1.1--2.9 \cite{haghshenasEnoughHotAir2023}  & Simple, low cost, widely adopted & Poor efficiency at high rack power ($>$30 kW) \\
Direct-to-Chip Liquid Cooling \cite{heydari2024experimental}  & High & $\sim$1.1--1.3\cite{Azarifar2024LiquidCooling} & Excellent thermal transfer, ideal for dense AI workloads & Requires plumbing and infrastructure; risk of leaks \\
Immersion Cooling \cite{kuncoro2019immersion}  & Very high ($>$50 kW/rack) & $\sim$1.02-1.04 \cite{haghshenasEnoughHotAir2023} & Ultra-efficient, low noise, compact & Redesign needed, fluid cost, specialized hardware \\
Hybrid Cooling \cite{heydari2024experimental} & Medium-high  & $\sim$1.34-1.38 \cite{zhouPowerUsageEffectiveness2022} & Flexible and adaptable to existing setups & System integration complexity \\
\bottomrule
\end{tabular}
\label{tab:cooling_comparison}
\end{table*}

\subsection{Other Supporting Facilities}

Beyond core IT hardware, electrical power infrastructure, and cooling systems, AI data centers incorporate additional supporting facilities that ensure reliable, secure, and efficient operations. These include: (\romannumeral1) \emph{physical infrastructure and layout} encompass data hall design features such as hot/cold aisle containment to optimize airflow, raised floors or overhead trays for organized cabling and ventilation, and structural reinforcements to accommodate the weight of high-density server racks; (\romannumeral2) \emph{monitoring and control systems} involve dense networks of temperature, humidity, and airflow sensors integrated into centralized building management platforms, along with fire detection and suppression systems employing inert gases or pre-action sprinklers to safeguard equipment; (\romannumeral3) \emph{security and access control} measures combine physical barriers, continuous video surveillance, and multi-factor authentication to prevent unauthorized entry; and (\romannumeral4) \emph{auxiliary systems} comprise redundant network links, battery energy storage systems, water treatment units for cooling operations, energy-efficient and emergency lighting, and dedicated office and maintenance spaces for on-site staff.

\section{Electricity Demand of AI Data Centers}
\label{sec:load}

This section analyzes the electricity demand of AI data centers. It first outlines their electricity consumption structure and efficiency, and then presents the characteristics and patterns of the dominant component, AI computing load, across different stages of AI workflows, including model preparation, training, fine-tuning, and inference.

\subsection{Electricity Consumption Structure and Efficiency}

AI data centers are emerging as one of the fastest-growing electricity consumers in the energy sector. 
According to the IEA report \cite{iea2024energyai}, global data centers consumed around 415 TWh of electricity in 2024, accounting for about 1.5\% of total global electricity consumption. Looking ahead, the IEA projects that global data center electricity demand will more than double by 2030, reaching around 945 TWh, with AI identified as the primary driver of this growth \cite{iea2024energyai}. The global data center electricity consumption and related project initiatives are further discussed in Appendix~\ref{app:global}.

Modern hyperscale AI facilities typically contract for power capacities exceeding 100 MW, with some campuses planned to scale to the gigawatt level, to support ultra-scale AI training and inference. 
According to the Electric Power Research Institute (EPRI) white paper \cite{aljbour2024powering}, electricity consumption in a data center is mainly attributed to IT hardware equipment, which typically accounts for 40-50\% of total load. Cooling systems represent the second-largest share, consuming approximately 30-40\%, depending on the cooling technology and server rack density. The remaining 10-30\% is from other supporting facilities, including lighting, office spaces, and monitoring and security systems. In AI data centers, the share of electricity consumed by IT equipment is higher and typically exceeds 60\%, due to the adoption of high-density GPU/TPU racks and efficient liquid cooling
systems. Nevertheless, the distribution of electricity consumption within a data center varies
substantially with facility design, climate, cooling technology, equipment
utilization, and operating conditions.

Power Usage Effectiveness (PUE) \cite{aljbour2024powering} is one of the most widely used metrics to measure the energy efficiency of a data center, which is defined as the ratio of total facility electricity consumption to IT-equipment electricity consumption
over the same accounting period:
\begin{align}\label{eq:pue}
  \text{PUE} = \frac{\text{Total Facility Electricity Consumption}}{\text{IT Equipment Electricity Consumption}} \geq 1.
\end{align}
A lower PUE indicates greater energy
efficiency, with a value of 1.0 signifying that all facility energy is delivered to IT equipment. 
A typical enterprise data center has a PUE of around 1.5-1.6, while 
modern large-scale facilities achieve a lower PUE below 1.3. For example, Google reports a comprehensive trailing twelve-month PUE of 1.09 across its large-scale data centers, with some sites operating under 1.06 \cite{googlePUE2025}. Similarly, 
Meta’s 2024 Sustainability Report states an average PUE of 1.08 across its data centers   \cite{metaSustainability2024}. These values indicate that IT equipment accounts for a dominant share of facility energy consumption and that facility-level energy overhead is relatively low.




\subsection{AI Computing Load Patterns at Different Stages}

Since AI computing constitutes the dominant share of electricity consumption in AI data centers, this subsection analyzes its load patterns and characteristics across the four main stages of the AI model lifecycle:
\textit{preparation}, \textit{training}, \textit{fine-tuning}, and \textit{inference} \cite{touvron2023llama}.

\subsubsection{Preparation Stage}
The preparation stage aims to prepare the AI model architecture and data before formal training begins. It includes: (\romannumeral1) \emph{model preparation}: designing or selecting a base AI model, configuring model size, parameters, architecture, precision settings, and conducting early-phase experimental testing; and (\romannumeral2) \emph{data acquisition and pre-processing:} LLMs, which contain billions to hundreds of billions of parameters, require massive training datasets (often
 terabytes in size) that cover a broad range of content and are collected from diverse sources such as Wikipedia, CommonCrawl \cite{commoncrawl}, BookCorpus \cite{zhu2015aligning}, Reddit posts, GitHub code, and other text datasets \cite{zhao2023survey}. 
These raw datasets must be cleaned and preprocessed into training-ready formats (e.g., tokenization, normalization, augmentation). The extract-transform-load pipelines often rely on distributed computing clusters that consume significant electricity during large-scale pre-processing.
Well-designed AI models and carefully processed data are critical to the model training efficiency.
Empirical results in \cite{verdecchia2022data} show that data-centric approaches, such as removing redundant samples from datasets or performing effective feature selection, can drastically
reduce the energy consumption of model training by more than 90\%, underscoring the importance of developing Green AI \cite{verdecchia2022data}.
Nevertheless, the preparation stage is flexible and widely dispersed, with the workloads varying substantially depending on factors such as the choice of data processing libraries and tools, the deployment environment, and the workflow structure, making its electricity consumption difficult to quantify precisely.

\subsubsection{Training Stage}

The training stage is the most electricity-intensive phase of AI development, particularly for LLMs. 
Training often runs continuously for days or weeks, during which computing hardware operates at near-peak capacity, resulting in a significant and sustained power draw. 
For example, training GPT-3 is estimated to have consumed 1.29 GWh of electricity \cite{patterson2021carbon};
this intensive process involved approximately 14.8 days of continuous computation on 10,000 NVIDIA V100 GPUs in a Microsoft data center \cite{patterson2021carbon}. 
Furthermore, training GPT-4 required an estimated over 50 GWh of electricity, approximately 40 times more than GPT-3, and equivalent to nearly 0.02\% of California's annual electricity consumption\footnote{OpenAI does not officially disclose the energy consumption data for training its GPT models. The figures cited are estimates based on third-party analysis and research.}~\cite{extnet_gpt4_energy}.
Similar LLMs, such as Meta’s LLaMA series and Google’s Gemini, require tens of thousands of high-performance GPUs operating in parallel for training, resulting in exceptionally high electricity consumption, often ranging from hundreds to thousands of MWh.

Typical electricity load patterns during the training stage include:  
(\romannumeral1) \emph{Initial load ramp-up phase}: at the beginning of training, resource utilization gradually increases as datasets are loaded into memory and distributed across computing nodes, creating a short period of escalating demand \cite{mughees2025forecast}.
(\romannumeral2) \emph{Constant high demand}: computing hardware operates at near-maximum utilization for prolonged periods with a very high power load throughout the training stage \cite{Li2025AILoadDynamics}.  
(\romannumeral3) \emph{Rapid and large power fluctuations}: the alternation between a power-intensive computation phase and a less power-demanding communication phase in the training stage often leads to large power swings \cite{choukse2025power}. In addition,
tasks such as checkpointing, intermediate result saving, and 
large-scale data transfers between nodes can cause brief but pronounced spikes in electricity consumption; 
additional transient surges may occur when training is paused and resumed \cite{mughees2025forecast,Latif2024EmpiricalGPU}. 
(\romannumeral4) \emph{High cooling load}: sustained full-power operation of GPUs and supporting equipment produces substantial heat, requiring cooling systems to operate at elevated capacity, which significantly increases total energy use. Electricity consumption during the training stage is affected by many factors, such as AI model architecture, training configuration choices, batch size, and numerical precision requirements. Several strategies have been proposed to improve training efficiency. 
For example, reference~\cite{McDonald2022May} showed that power-capping, which limits the maximum power a GPU can consume, enables a 15\% decrease in energy usage with a marginal increase in overall computation time. Other approaches include dynamically adjusting training parameters based on convergence progress and implementing early stopping when performance improvements plateau.

\subsubsection{Fine-tuning Stage}

The fine-tuning stage adapts a pre-trained AI model to a specific downstream task, such as legal text analysis, medical report summarization, or customer service chatbots. 
Since the AI model has already undergone extensive pre-training, fine-tuning typically requires substantially less computation and electricity than training from scratch. Empirical measurements in \cite{Wang2023FineTuningEnergy} show that pre-training BERT consumes electricity equivalent to approximately 400 to 45,000 fine-tuning runs, depending on the dataset size and task complexity.
Nevertheless, as fine-tuning often involves smaller datasets, frequent evaluation cycles, and hyperparameter tuning, power use tends to be more variable. Rather than maintaining a sustained high baseline, fine-tuning workloads typically exhibit intermittent bursts of GPU utilization followed by lower activity periods. 

Typical electricity use patterns during fine-tuning include: (\romannumeral1) \emph{moderate average demand}, generally lower than full pre-training but varying by model size and dataset; (\romannumeral2) \emph{bursty workloads}, with short periods of high consumption during forward/backward passes, validation, and hyperparameter exploration; and (\romannumeral3) \emph{cooling demand variability}, with cooling loads fluctuating in step with computational bursts. Although a single fine-tuning run consumes far less energy than pre-training, its much higher frequency across numerous organizations implies that fine-tuning can cumulatively represent a non-negligible share of total lifecycle energy consumption, underscoring the need to enhance its efficiency. For example, reference
\cite{Huang2023GreenTrainer} introduced GreenTrainer, an adaptive fine-tuning algorithm that strategically selects which model parameters to update. This method reduces fine-tuning workload by up to 64\% without significant accuracy loss, highlighting the potential to make fine-tuning more energy-efficient \cite{Huang2023GreenTrainer}. Reference ~\cite{Rehman2025Jan} analyzes the trade-offs between energy consumption and model performance, showing that
fine-tuning smaller models such as T5-base, BART-base, or LLaMA3-8B can achieve competitive results while significantly reducing energy usage.

\subsubsection{Inference Stage}

Inference is the process of executing a well-trained AI model to generate outputs in response to user inputs. On a per-query basis, it is the least electricity-intensive stage of the AI model lifecycle. Nevertheless, as inference is performed continuously at scale, often serving millions or billions of requests, its cumulative electricity consumption can exceed that of training. According to
\cite{patterson2022carbon}, compared with 40\% for training, inference accounts for approximately 60\% of total AI energy usage at Google, owing to the  billion-user services that rely on AI. Recent estimates suggest inference can account for up to 90\% of a model’s total lifecycle
energy use \cite{jegham2025hungry}. 
The electricity consumption for a single AI inference query varies widely depending on the model size and architecture, prompt length, and task complexity \cite{jegham2025hungry}. 
Compared to a traditional Google search, which consumes about 0.3~Wh  \cite{kanoppi2024searchenergy}, 
a recent study \cite{jegham2025hungry} estimates that OpenAI o3 consumes 39.2~Wh, DeepSeek-R1 33.6~Wh, and GPT-4.5 30.5~Wh for processing a long prompt query, whereas GPT-4.1 Nano requires only 0.45~Wh. Reference \cite{Luccioni2023Nov} compares inference energy consumption across a wide range of tasks and shows that image generation is among the most energy-intensive, whereas simpler tasks, such as text classification, consume significantly less electricity.

Unlike the sustained, high load profile of model training, multi-modal generative AI inference workloads are characterized by significant variability. This is driven by 
irregular user request arrivals, variable prompt lengths, diverse query complexities, and the unpredictable nature of real-time interactions, resulting in short, intense bursts of computing load. As a result, typical electricity use patterns in the inference stage include:  
(\romannumeral1) low average demand per query but high total demand over time due to continuous operation;  
(\romannumeral2) bursty and unpredictable spikes from user activity and varying task complexity; and 
(\romannumeral3) strong diurnal patterns, with peak usage during business or social activity hours.  
For example, BurstGPT provides 5.29 million traces from regional Azure OpenAI GPT services over 121 days and reveals diversified burstiness and concurrency patterns across services and model types~\cite{Wang2024BurstGPT}.
Supporting such dynamic loads requires scalable and flexible IT and power infrastructure capable of rapidly adjusting resources to provide fast response and maintain operational reliability.
Energy efficiency can be improved at both the hardware and software levels. Specialized low-power accelerators such as FPGAs and ASICs  can reduce energy consumption, while techniques like model quantization, pruning, and knowledge distillation further enhance performance~\cite{ Stock2020Distillation}. Emerging architectures such as retrieval-augmented generation (RAG) \cite{fan2024survey} introduce additional complexity by combining model inference with large-scale data retrieval processes. Despite the exponential growth in model parameters, the associated energy cost of inference has been shown to grow sub-linearly, suggesting that efficiency gains can partially offset the demands of larger architectures~\cite{Desislavov2021Sep}.

\begin{figure}[ht]
    \centering
    \includegraphics[width=1\linewidth]{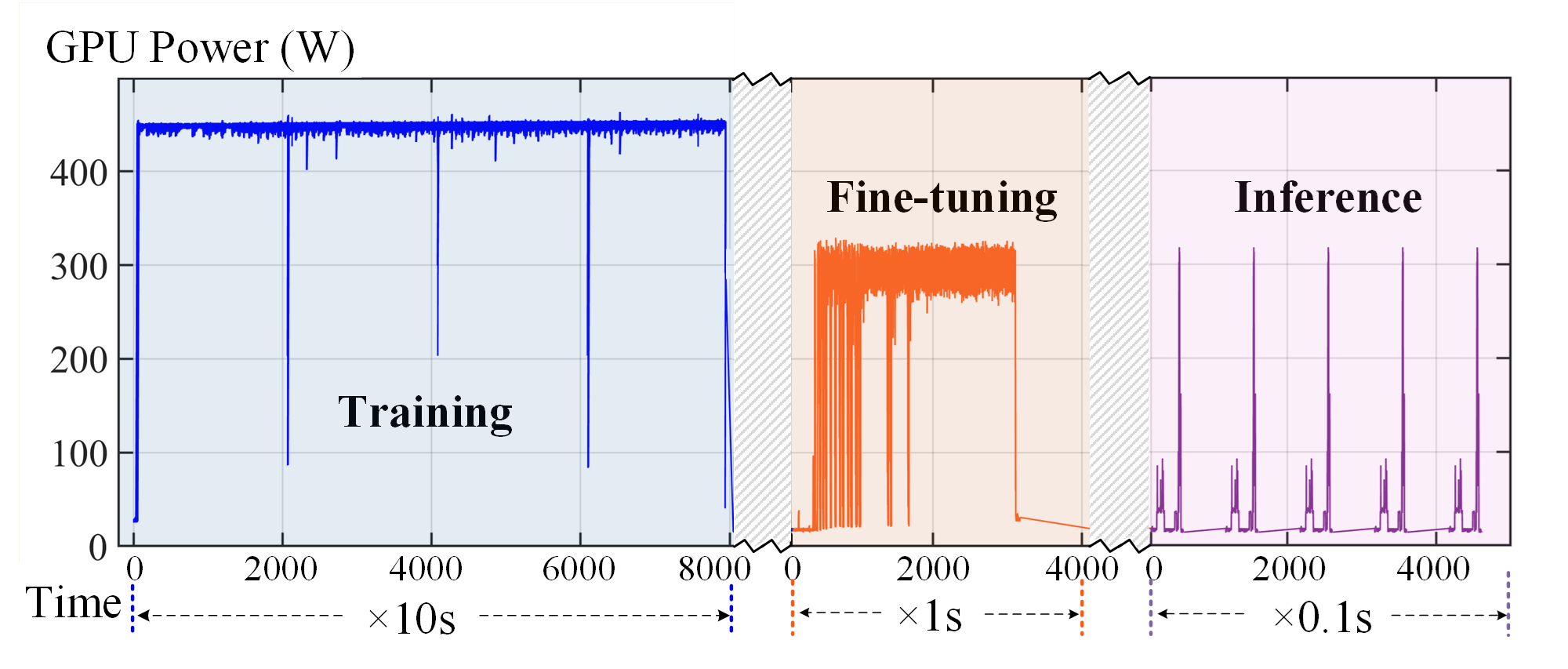}
    \caption{Illustration of the patterns of AI computing load (represented by GPU power) during training, fine-tuning, and inference stages. (The data are derived from~\cite{Li2024Sep}; refer to Figures 9, 13, 15, and Table III in~\cite{Li2024Sep} for detailed data and experimental settings.) }
    \label{fig:load_pattern}
\end{figure}

\begin{figure*}[ht]
    \centering
    \includegraphics[width=0.8\linewidth]{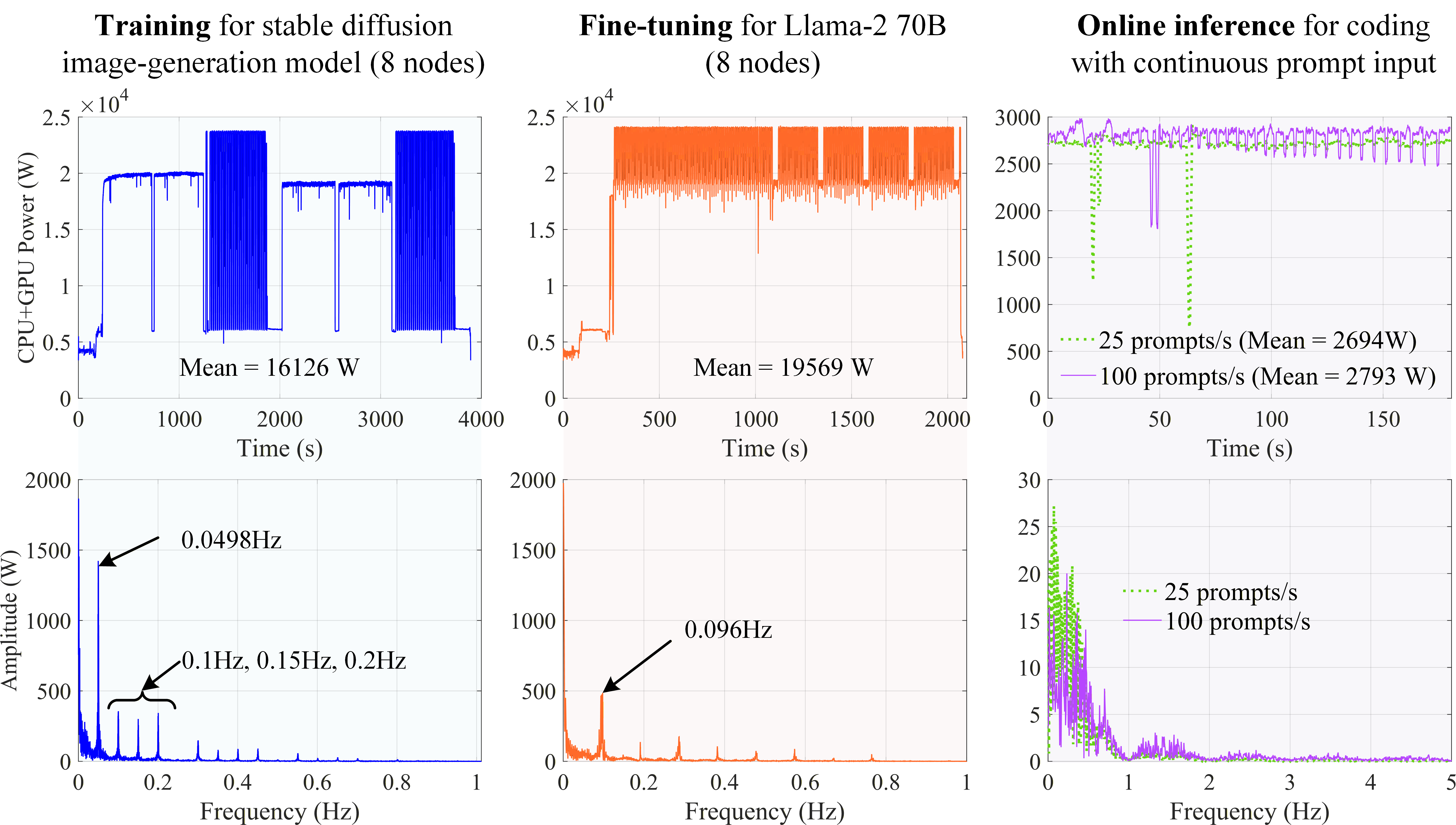}
    \caption{Illustration of the time-domain and frequency-domain characteristics of AI computing loads, represented by CPU and GPU power consumption, during the training, fine-tuning, and inference stages, based on the open-source data from \cite{Vercellino2026GenAIPowerProfiles}. (The upper three plots show the time-domain power profiles of training, fine-tuning, and online inference workloads, while the lower three plots show the corresponding frequency-domain profiles obtained using the fast Fourier transform (FFT). The mean power value for each stage is presented as ``Mean = '' in the time-domain plots, which is removed in the FFT results. For the online inference workload, the legends indicate different prompt request rates; more experimental details are provided in~\cite{Vercellino2026GenAIPowerProfiles}.)} 
    \label{fig:ThreestagesNRL}
\end{figure*}

\subsection{Available Datasets, Summary, and Discussion} \label{sec:discussion}

AI computing load is highly energy-intensive, with each stage of the workflow contributing differently to the total electricity consumption. According to  \cite{aljbour2024powering}, the inference stage represents a dominant share of approximately 60\% of the energy footprint, while training accounts for 30\%, and model preparation and fine-tuning make up the remaining 10\%. 
 Moreover, these stages exhibit distinct load patterns.
Several publicly available datasets and measurement studies provide useful resources for studying AI computing power profiles. These include the Dataset of Generative AI Workload Power Profiles from the National Laboratory of the Rockies (NLR)~\cite{Vercellino2026GenAIPowerProfiles,NLR2026GenAIPowerProfiles}, Watt Counts~\cite{Argerich2026WattCounts} (an open-access energy dataset and benchmark for LLM inference), and high-performance computing datasets with power-monitoring information, such as the MIT Supercloud Dataset~\cite{supercloud}, M100~\cite{Borghesi2023M100}, and PM100~\cite{Antici2023PM100}. Nevertheless, these existing datasets primarily consist of laboratory-scale experiments and GPU-level AI workload and power measurements, rather than full-scale operational data from commercial AI data centers. Developing anonymized, industry-supported operational datasets for AI data centers remains an important direction for future research and industry collaboration.

Figures~\ref{fig:load_pattern} and~\ref{fig:ThreestagesNRL} illustrate the time-varying electric load patterns across different stages of AI computing based on real experimental measurements. The electricity loads for each stage in Figure \ref{fig:load_pattern} are derived from \cite{Li2024Sep}, where the workloads were executed using GPT-2. Figure~\ref{fig:ThreestagesNRL} presents the raw power measurement data of AI computing from~\cite{Vercellino2026GenAIPowerProfiles,NLR2026GenAIPowerProfiles}, and we further apply the fast Fourier transform to convert the time-domain measurements into the frequency domain and analyze their frequency-domain characteristics and dominant spectral peaks. Detailed experimental settings are provided in~\cite{Li2024Sep,Vercellino2026GenAIPowerProfiles}. As shown in Figures \ref{fig:load_pattern} and \ref{fig:ThreestagesNRL}, an initial rapid ramp-up phase is observed, during which power consumption rises sharply from a low (near-zero) value to a high operational level. In the training stage, power remains consistently high, characterized by intermittent large rapid load fluctuations, along with persistent high-frequency variations, the magnitude of which depends on the underlying hardware and training algorithms \cite{Li2024Sep}. The fine-tuning stage begins with pronounced high-frequency, large-magnitude fluctuations, which gradually decrease in frequency over time; however, moderate fluctuations continue throughout the process. 
Power fluctuations during training and fine-tuning arise from both workload-level and hardware-level dynamics, including alternating computation and communication phases, GPU utilization changes, checkpointing or data movement, batching behavior, and power-management control of accelerators.
In contrast, the inference stage operates over shorter durations and exhibits highly variable short bursts of power demand, driven by the stochastic nature of real-time inference requests. This diversity in load profiles underscores the critical need for advanced AI data center designs, grid integration strategies, and power-aware workload scheduling approaches that balance performance, sustainability, and system reliability.


\section{Grid Impacts and Emerging Challenges of AI Data Centers}
\label{sec:gridimpact}

With immense electricity consumption and large, rapid power fluctuations, large-scale AI data centers introduce emerging impacts and operational challenges for power grids. This section analyzes these challenges across three interrelated timescales: \emph{long-term} planning and interconnection, \emph{short-term} operation and electricity markets, and \emph{real-time} grid dynamics and stability. This multi-timescale perspective illustrates how AI data center loads affect different stages of power grid management. Other critical cross-cutting challenges, including cybersecurity, decarbonization, and water consumption, are discussed as well.

\subsection{Long-Term Planning and Interconnection}

\begin{figure}[ht]
\centering
\includegraphics[width=1\linewidth]{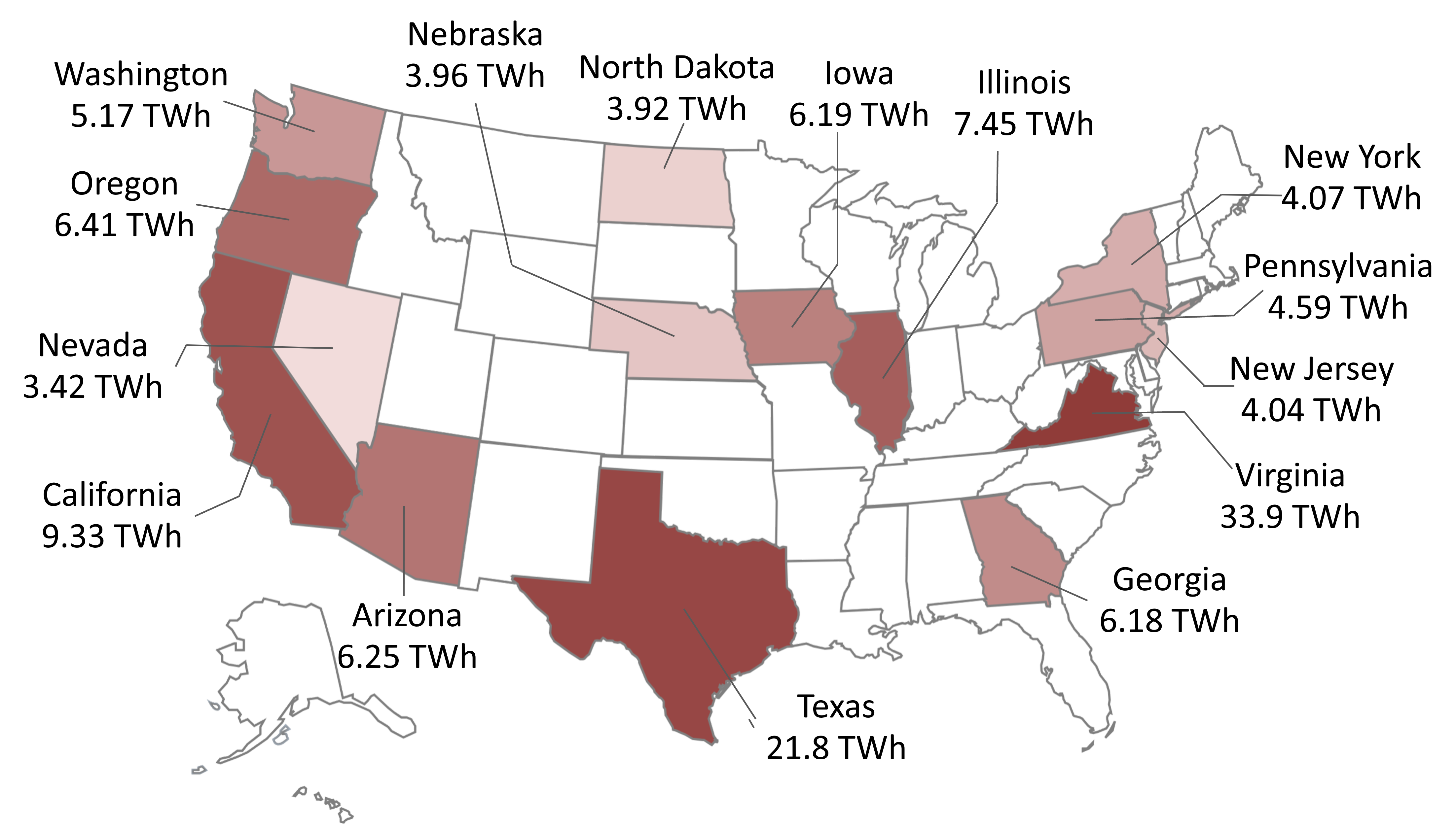}
\caption{The fifteen U.S. states with the highest data center loads in 2023 \cite{aljbour2024powering} (darker colors indicate higher loads).}
\label{fig:AI Data Center Load by Region}
\end{figure}

Over the long-term timescales (years to decades), 
AI data centers are projected to drive a sustained increase in electricity demand, requiring substantial investment in generation capacity, transmission infrastructure, and distribution equipment to maintain an adequate and reliable electricity supply. 
In practice, many regional grids are incapable of accommodating large-scale data centers without extensive transmission and distribution upgrades, which often require 5-10 years for planning, permitting, and construction~\cite{lin2024exploding}. Reference~\cite{lin2024exploding} assessed the 5‑year projections of AI data-center demand across several bulk power grids and found that resource‑adequacy constraints could severely limit their planned growth, particularly in high‑density clusters such as EirGrid in Ireland and Dominion in the U.S. Moreover, AI data centers tend to concentrate in a limited number of regions, driven by low energy prices, available land and water resources, and supportive policy environments. For example, in the U.S., fifteen states (notably Virginia, Texas, and California) accounted for an estimated 80\% of the
national data center load in 2023 \cite{aljbour2024powering}, as shown in Figure \ref{fig:AI Data Center Load by Region}. This geographical concentration effect further intensifies regional grid stress. 

To address this challenge,
\emph{coordinated co-planning} of AI data centers, the power grid, and generations can reduce costly infrastructure investments, alleviate transmission congestion, 
enhance operational reliability, and support decarbonization goals. This approach integrates transmission and distribution expansion planning with the strategic siting of AI data centers, as well as the optimal  siting and sizing of additional generation and energy storage resources. A number of studies  \cite{chen2024joint,guo2021integrated,vafamehr2017framework} focus on the modeling and optimization of joint planning for conventional data centers and the power grid, while solutions tailored to AI data centers remain underexplored. Mathematically,
coordinated co-planning of AI data centers and grid infrastructure can be formulated as a multi-period optimization problem that jointly determines grid expansion, data center siting and capacity, generation and storage investment, and flexible load commitments~\cite{guo2021integrated,11341378}. The objective is to minimize total investment and multi-period operational costs over the planning horizon, subject to power flow constraints, generation and transmission capacity limits, resource adequacy requirements, interconnection constraints, data center service-quality constraints, carbon and water limits, investment budget constraints, and other operational or regulatory requirements. In practice, this co-planning problem is further complicated by uncertainty in future AI workload growth, renewable generation, electricity prices, fuel availability, interconnection timelines, and permitting delays. Therefore, stochastic optimization~\cite{chen2024joint}, robust optimization~\cite{11355926}, or distributionally robust optimization~\cite{11341378} formulations can be used to evaluate planning decisions under various future scenario uncertainties. Such frameworks provide a quantitative basis for comparing tradeoffs among investment cost, reliability, carbon emissions, water consumption, and deployment speed.

Another key strategy is \emph{co‑location with dedicated local generation}, such as renewable generators and nuclear power plants. This scheme can directly supply large, stable power sources to high-demand AI workloads, reducing reliance on long-distance electricity transmission and mitigating grid-integration challenges. For example, Amazon Web Services (AWS) has arranged to source approximately 960 MW directly from a nuclear power plant in Pennsylvania to power its data center under a dedicated power purchase agreement \cite{aws2024}. Google and Kairos Power have signed a Master Plant Development Agreement to deploy a fleet of advanced small modular reactors to provide up to 500 MW of clean electricity to Google's AI data centers \cite{patel2024_google_kairos}.

In addition, the grid \emph{interconnection policies and regulatory frameworks} remain among the most significant bottlenecks to AI data center  deployment. For example, in 2025, Texas passed Senate Bill 6 (SB-6) \cite{TXSB6_2025}, which redefines the interconnection process for large electrical loads, such as data centers exceeding 75 MW, within the Electric Reliability Council of Texas (ERCOT) grid. Under this bill, large load customers must submit interconnection requests through the interconnecting utility and comply with new requirements, including disclosure of similar requests, reporting of on-site backup generation, payment of a minimum \$100k transmission screening study fee, proof of site control, and a financial commitment to cover transmission infrastructure costs. Complex regulatory and permitting workflows, thorough grid integration studies, overloaded interconnection queues, and the need for major grid infrastructure upgrades are causing multiyear delays in data center deployments across regions such as Northern Virginia \cite{FTI2024PowerTrends}. 
Refining interconnection processes and deploying AI and automation tools to accelerate interconnection studies could significantly reduce these delays. 
For example, ERCOT's new Batch Zero framework~\cite{ERCOT2026BatchConnection} replaces previous project-by-project studies with coordinated batch studies, aiming to streamline the grid interconnection study process for large loads. 
Moreover, \emph{tariff structure} varies across jurisdictions: many data centers fall under smaller regional utilities and are subject to distribution-level tariffs, while hyperscale facilities connect directly to bulk transmission systems, exposing them to wholesale market price signals. As a result, the tariff design and standardization are critical to ensuring cost predictability and sustaining the deployment of AI data centers.

\subsection{Short-Term Operation and Electricity Markets}

On the short-term timescales (hourly to weekly), the large-scale integration of AI data center loads (ranging from tens to hundreds of megawatts) introduces substantial operational and market challenges to the power grid. 

\subsubsection{Impacts on Power Dispatch and Reserve Scheduling}

As economically available energy storage remains limited relative to the scale and duration of power system needs, maintaining the power balance between generation and demand remains a fundamental objective of power grid operation. On the short‑term timescale, power balance is achieved through {unit commitment} (UC) and {economic dispatch} (ED), supported by
 {reserve scheduling} to accommodate unforeseen power variations~\cite{wood2013power}. 
The immense and bursty power usage of AI data centers, particularly during intensive model training and large-scale inference, introduces significant uncertainty and rapid load variations to grid operation~\cite{mughees2025forecast}. This challenges conventional deterministic power dispatch schemes and necessitates advanced short‑term forecasting techniques for AI data center loads~\cite{mughees2025forecast,li2023shortterm}. However, accurate prediction remains inherently difficult due to the complex, non‑cyclic patterns and dynamic scheduling behaviors of AI workloads. In addition, the rapid and unpredictable ramping of AI data center loads imposes greater reserve requirements. To ensure reliability amid large and fast load variations, grid operators may need to maintain elevated spinning and non‑spinning reserve margins, which increase operational costs and call for additional fast‑responding reserve resources.

\subsubsection{Impacts on Electricity Markets and Prices}

Large-scale AI data center loads are expected to 
 place upward pressure on \emph{capacity market clearing prices}, as higher peak demands require additional capacity commitments. This impact has 
been observed in PJM, where capacity market clearing prices for the 2026-2027 delivery year increased to \$329.17/MW-day, over ten times higher than the price of \$28.92/MW-day in the 2024-2025 delivery year, with rapid data center growth identified as a major contributing factor~\cite{pjm2025capacity,kunkel2025pjm}.
Similar effects are emerging in New York ISO, where anticipated data center expansion is influencing forward capacity markets and resource adequacy planning~\cite{nyiso2025powertrends}. 
In addition, the substantial and often unpredictable demand from AI data centers raises marginal generation costs in wholesale markets, as more expensive mid-merit and peaking units are dispatched to meet AI-driven loads~\cite{aljbour2024powering}. It may also amplify price volatility in day-ahead and real-time markets due to short-term large AI load fluctuations. 
Rapid AI load surges can result in sharp increases in locational marginal prices (LMPs), particularly in transmission-congested areas \cite{JLARC2024DataCenter}.

These impacts highlight a broader challenge in electricity market design: how to accommodate rapidly growing AI data center loads while allocating the associated costs fairly and efficiently. As large AI data centers can drive new capacity procurement, transmission expansion, reserve requirements, and congestion-management costs, future market rules and tariffs should better reflect cost-causation principles. In particular, system operators and regulators may need to distinguish between firm, inflexible loads that require guaranteed supply during peak conditions and flexible or interruptible AI loads that can reduce demand, shift workloads, or use on-site resources during system stress. Such differentiation can help ensure that data centers contributing to peak demand and infrastructure expansion bear an appropriate share of the associated costs, while those providing verified flexibility are properly compensated. 
Meanwhile, cost-allocation mechanisms should avoid broadly socializing AI infrastructure costs to existing customers who do not directly benefit from or cause the new demand growth. Therefore, designing market mechanisms that preserve resource adequacy, incentivize grid-supportive operation, and fairly allocate price increases remains an important open problem for system operators, regulators, and policymakers.

\subsection{Real-Time Dynamics and Grid Stability}
\label{sebsec:Real-Time Dynamics and Grid Stability}

In real-time and fast timescales (sub-seconds to minutes), large-scale AI data center loads can significantly impact power system dynamics and stability. These facilities are inherently power electronics-based loads, with their dynamic behavior governed by the control of internal power electronic converters and by the AI workload dynamics.

\subsubsection{Grid Disturbance Ride-Through} \label{sec:ridethrough}

\begin{figure}[ht]
\centering
\includegraphics[width=1\linewidth]{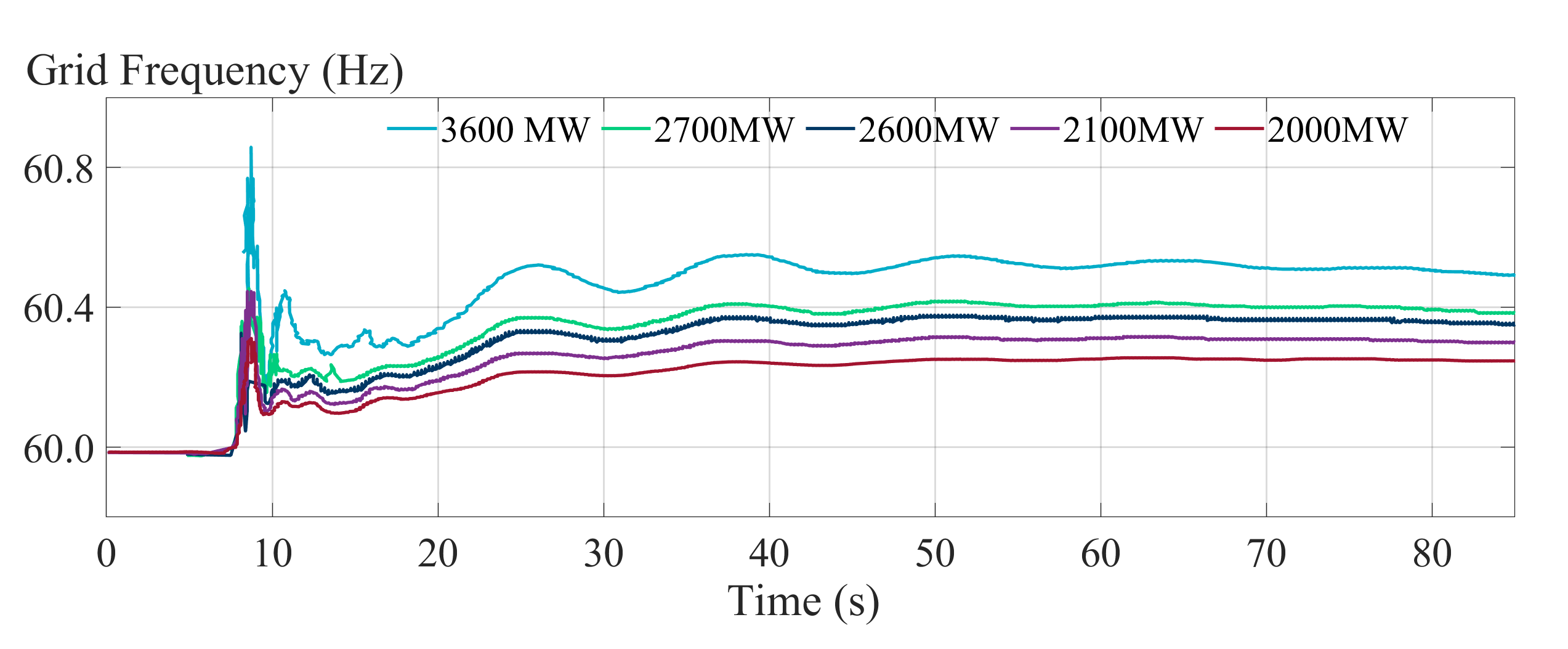}
\caption{ERCOT system frequency response to several large electronic load trip events \cite{Jeff2025_ERCO}. (The results show that sudden disconnection of large electronic loads can cause significant frequency overshoot, with larger tripped load levels leading to more severe frequency excursions. This indicates that insufficient ride-through capability of large AI data center loads could exacerbate grid disturbances and increase the risk of cascading events.)}
\label{fig:LargeLoadLoss}
\end{figure}

AI data centers interface with the power grid through high-capacity power electronic converters, which are sensitive to short-duration voltage and frequency disturbances at the point of interconnection, such as voltage sags, voltage swells, and upward or downward frequency deviations. 
These disturbances can induce substantial power oscillations and even trigger the disconnection of an AI data center from the grid to protect internal electronic devices.
 Without adequate fault ride-through (FRT) capabilities or grid-disturbance
tolerance, such sudden trips can occur during grid faults and potentially
exacerbate the original disturbance. In regions with high concentrations
of AI data centers and other large electronic loads, a common grid
disturbance could cause multiple facilities to trip nearly simultaneously.
The resulting sudden loss of load can produce significant frequency and
voltage excursions and may trigger additional load or generator trips,
thereby increasing the risk of cascading outages
~\cite{billo2025large_load_ridethrough}.
 These impacts have been observed in practice. For example, 
 high-frequency undamped power oscillations appeared in a data center-rich region in Dominion Energy's grid, triggered by a once-a-second voltage sag \cite{mishra2025understanding}. 
 ERCOT has also reported multiple events wherein  large computing loads tripped during transmission faults due to  insufficient ride-through capability \cite{Gravois2024_ERCO}. For example, Figure \ref{fig:LargeLoadLoss} illustrates the grid frequency dynamics under several large electronic load trip events studied in ERCOT \cite{Jeff2025_ERCO}. It shows that a loss of 2700 MW of such loads can drive system frequency up to 60.4 Hz, a level at which conventional generators may fail to ride through, potentially leading to uncontrolled cascading events \cite{Jeff2025_ERCO}. To this end, ERCOT considered imposing ride-through requirements for large data center and crypto-mining loads \cite{ERCOT2025_MB062325} to ensure reliable interconnection and grid operation.

\begin{figure}[htbp]
    \centering
    \includegraphics[width=1\linewidth]{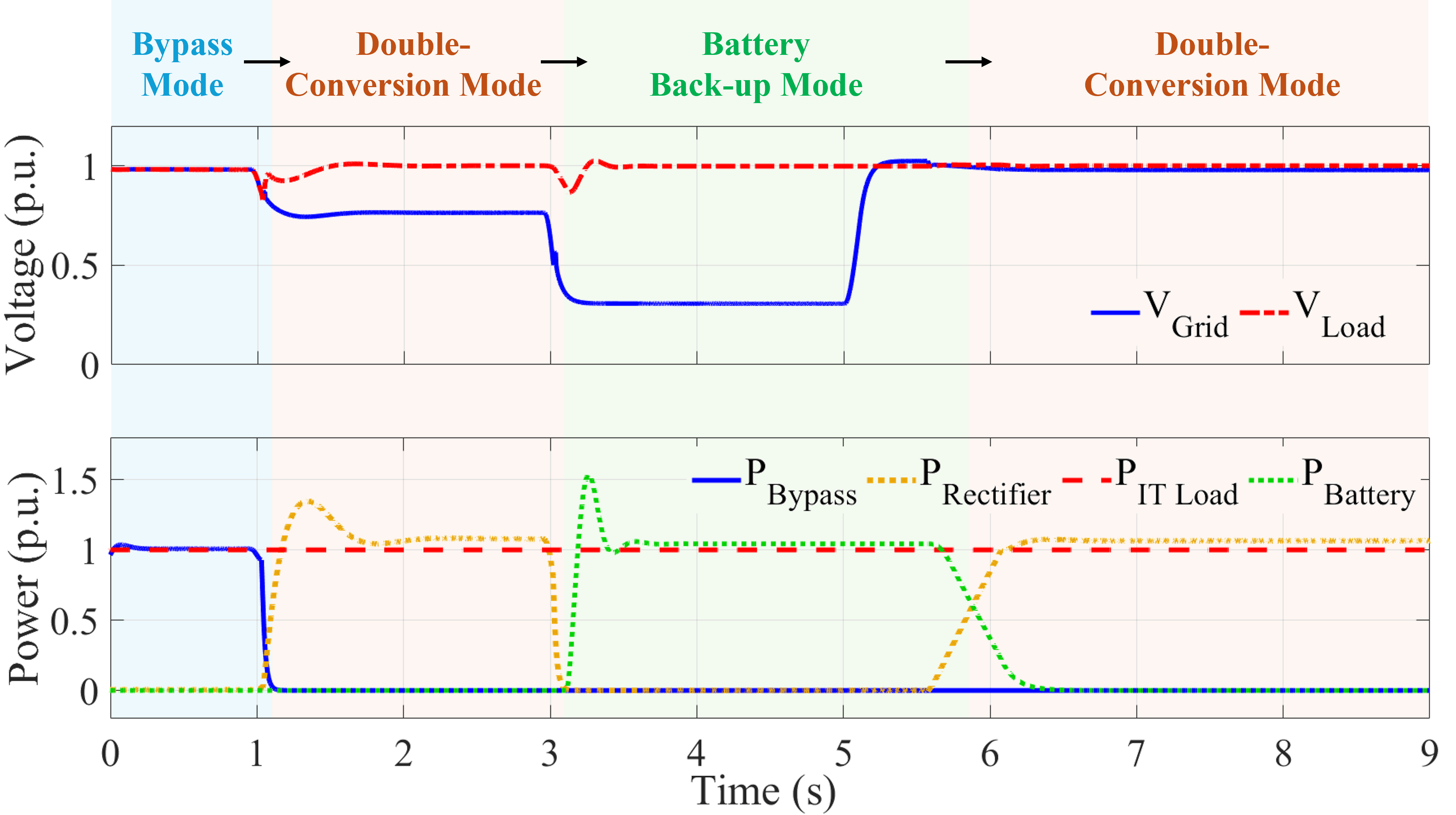}
    \caption{Simulated operating mode transitions and reconnection behavior of an AI data center during voltage sag and recovery. 
(The simulations are conducted in PSCAD~\cite{pscad_manual} using the open-source EMT models of AI data center loads developed in~\cite{Wang2026AIDCPSCAD}.  As shown in Figure~\ref{fig:threemode}, a constant IT load power profile is simulated for illustration and the cooling load is not included.
   The upper subplot shows the trajectories of the grid-side voltage magnitude, ${V}_{\text{Grid}}$, and the IT load-side voltage magnitude, ${V}_{\text{Load}}$. The lower subplot shows the active-power trajectories through the bypass switch, ${P}_{\text{Bypass}}$, through the double-conversion rectifier, ${P}_{\text{Rectifier}}$, the IT load power, ${P}_{\text{IT Load}}$, and the UPS battery output power, ${P}_{\text{Battery}}$. 
 From 0 to 1~s, the grid voltage is approximately 1~pu and the data center operates in bypass mode, with the IT load supplied directly by the grid. At 1~s, a moderate voltage sag to 0.75 pu triggers double-conversion mode, where the rectifier-inverter interface maintains the load voltage near 1~pu. At 3~s, a severe voltage sag to about 0.3~pu causes the UPS to switch to battery backup mode, with the battery supplying the IT load. After the grid voltage recovers at 5~s and remains stable for a specified reconnection delay, the UPS reconnects to the grid-side supply and returns to double-conversion operation.)}
    \label{fig:modeswitch}
\end{figure}

As AI data centers rely on rectifier-based power electronic interfaces to convert grid-side AC power into DC power for downstream IT equipment, their internal DC computing loads, such as AI racks, are partially decoupled from grid frequency variations, and thus many converter-fed IT loads exhibit reduced intrinsic frequency sensitivity. In contrast, voltage disturbances, particularly voltage sags, can more directly affect AI data centers. Intuitively, for a given computing load power, $P=VI$, a reduction in grid-side input voltage $V$ requires the rectifier or upstream converter to draw higher current $I$, which can more easily trigger converter current limits and protection actions. Such voltage-sag disturbances may cause UPS mode transitions or, in severe cases, load disconnection at the point of interconnection. As a result, low-voltage ride-through (LVRT) capability is critical for AI data centers and other large loads in practice. 
Figure~\ref{fig:modeswitch} illustrates the simulated\footnote{The simulations are conducted in PSCAD~\cite{pscad_manual} using the high-fidelity, open-source electromagnetic transient (EMT) models of large AI data center loads developed by Texas A\&M University and ERCOT~\cite{Wang2026AIDCPSCAD}.} data center operating-mode switching process from bypass mode to double-conversion mode and then to battery backup mode under sustained voltage sags, as well as the subsequent reconnection process back to double-conversion mode after voltage recovery.

\subsubsection{Power Fluctuations and Grid Stability}

AI data centers can introduce rapid power fluctuations during cold starts, planned shutdowns, workload transitions, or sudden load interruptions triggered by faults or operational contingencies. For example, training large-scale AI models across tens of thousands of GPUs can induce significant power swings, arising from the synchronous nature of these jobs and the alternation between a power-intensive computation phase and a less power-demanding communication phase \cite{choukse2025power}. In addition, 
inference workloads often exhibit rapid ``start-stop'' behavior with sharp spikes followed by sudden drops in power consumption, as shown in Figures \ref{fig:load_pattern} and  \ref{fig:ThreestagesNRL}. 
Unlike conventional electromechanical loads, power electronics-based AI data center demand can change significantly within (sub-)second timescales, posing substantial stability risks to the grid.

\begin{figure*}[htbp]
    \centering
      \includegraphics[width=0.85\linewidth]{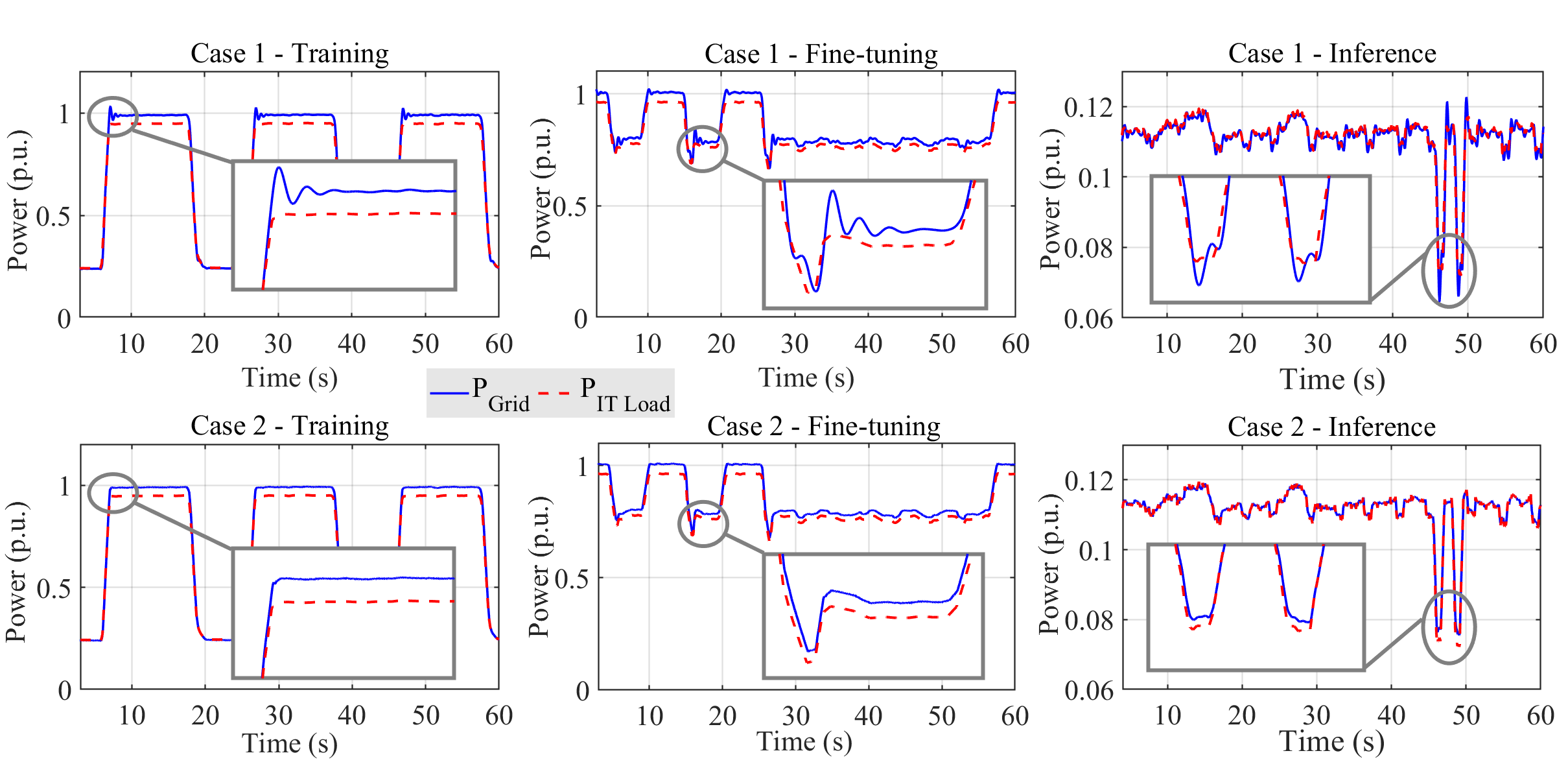}
    \caption{Comparison between device-level IT load power and facility-level grid-side power during the training, fine-tuning, and inference stages based on the electromagnetic transient (EMT) simulations in PSCAD. (The IT load profiles are obtained from one-minute segments of the open-source dataset reported in~\cite{Vercellino2026GenAIPowerProfiles}, which is the same dataset used in Figure~\ref{fig:ThreestagesNRL}. Two cases with different rectifier controller parameters are presented: Case~1 with lower damping, corresponding to the upper three subplots, and Case~2 with higher damping, corresponding to the lower three subplots.)}
    \label{fig:ThreestagesPSCADSim}
\end{figure*}


Such abrupt power variations can challenge the grid's frequency stability, induce voltage excursions, and interact with the dynamics of grid-connected generation and control systems, potentially exciting  oscillations and instability. For instance, a major resonance event was reported in 2017 across multiple Facebook data centers and analyzed in~\cite{LowFrequencyModel}. As noted in~\cite{ImpedanceModel}, unlike solar plants and wind farms, data centers typically have internal impedances much higher than those of transmission lines or distribution transformers,
which can amplify such resonance
phenomena. From a data center perspective,
reference \cite{choukse2025power} proposes three classes of power stabilization techniques to mitigate large power swings, including  (\romannumeral1) software-based methods that inject controlled workloads to smooth power transitions, (\romannumeral2) GPU-level firmware mechanisms that enforce ramping constraints and power floors, and (\romannumeral3) rack-level energy storage (e.g., supercapacitor) systems to reduce fluctuations by
absorbing and releasing power as needed.
Beyond these approaches, addressing the challenge of large power fluctuations also requires accurate dynamic modeling of AI data center behavior and close coordination between grid operators and data center operators to ensure effective mitigation and maintain system-level stability. 
Moreover, as data centers typically employ fast-acting UPS systems for power continuity, transitions between grid supply and UPS operation during disturbances must be carefully managed to prevent large transient power swings or synchronization issues upon reconnection.

In Section \ref{sec:discussion},  Figure~\ref{fig:ThreestagesNRL} presents real measured GPU/device-level load profiles and power fluctuations, whereas publicly available real facility-level AI data center load profiles remain lacking. To bridge the gap between device-level IT load power and facility-level grid-side power, we performed high-fidelity EMT simulations in PSCAD using the open-source AI data center model developed in~\cite{Wang2026AIDCPSCAD} to investigate their relationship. Specifically, the data center with UPS systems operates in double-conversion mode, as shown in Figure~\ref{fig:threemode}, and the cooling load is not included. 
We use the measured GPU power trajectories in Figure~\ref{fig:ThreestagesNRL} for the training, fine-tuning, and inference stages as the IT load power profiles, $P_\text{IT Load}$, and obtain the corresponding grid-side power profile, $P_\mathrm{Grid}$, for each stage through EMT simulations. Since the dynamic response depends on power electronic converter control, we consider two representative cases by adjusting the rectifier outer-loop voltage control parameters: Case~1 with low damping, and Case~2 with high damping. The simulation results of these two cases are shown in Figure~\ref{fig:ThreestagesPSCADSim}. It is observed that the grid-side power closely follows the IT load power, with the steady-state difference mainly due to the power conversion losses between the grid and the IT servers. During transient periods, relatively slow IT load variations, such as those in the training and fine-tuning stages, are accurately tracked by the grid-side power. In contrast, faster load fluctuations, such as those in the inference stage, can exhibit different grid-side behaviors depending on both the fluctuation pattern and converter control parameters. Moreover, a low-damping rectifier controller, as in Case~1, may amplify rapid power variations and produce transient overshoot, whereas a high-damping controller, as in Case~2, tends to smooth fluctuations before they propagate to the grid. 

Generally, higher damping can reduce oscillations and transient overshoot, but it may also lower the effective response speed or bandwidth of the controller, leading to slower tracking of disturbances and load variations. Therefore, converter control plays a critical role in shaping the dynamic behavior of AI data centers. In addition, the dynamic response is affected by DC-link capacitance. A larger DC-link capacitor can provide short-term energy buffering, helping maintain the DC-link voltage and smooth power fluctuations, but it also increases cost and physical size. The aggregate grid-side response further depends on whether a large number of IT racks operate in a synchronized or asynchronous manner. Further studies are needed to better understand how AI data center design affects facility-level dynamic behavior.

\subsubsection{Power Quality Issues}
High‑frequency switching devices and power electronic equipment in data centers can introduce significant harmonic distortions to the grid, degrading local power quality
and posing risks to nearby electrical facilities. These distortions originate primarily from non‑linear power supplies, AC/DC conversions, and the control  of server PFC circuits~\cite{ieee519}. 
 In large-scale AI facilities, where thousands of server racks operate simultaneously, these harmonics can accumulate and propagate into the local distribution system, presenting risks to industrial and residential equipment, such as malfunctions, accelerated wear, overheating, and electrical fires.
According to a 2024 Bloomberg analysis~\cite{bloomberg2024ai}, sensor data from more than 700,000 homes reveal a strong correlation between proximity to data centers and declining power quality, with evident effects observed within 20 miles of major data center clusters. 
Beyond harmonic distortion, AI data centers may contribute to other power quality issues, such as 
voltage flicker during rapid load changes, unbalanced loads across three phases \cite{bollen_power_quality_2000}, and interharmonics caused by certain switching patterns in high‑frequency converters. These power quality issues can propagate through the grid, potentially degrading protection system performance, reducing metering accuracy, shortening equipment lifespans, and impairing control system stability~\cite{bollen_power_quality_2000}.
To mitigate power quality issues and maintain compliance with industry standards such as IEEE Standard 519‑2014~\cite{ieee519}, key approaches include: (\romannumeral1) active harmonic filtering at the AI data center facilities, (\romannumeral2) phase balancing and load redistribution to minimize neutral current distortion, (\romannumeral3) coordination with grid operators for harmonic monitoring and mitigation planning, and (\romannumeral4) UPS and converter design optimization to minimize harmonic generation at the source. 
As AI data center capacity keeps expanding, proactive power quality management will be critical for maintaining grid stability and ensuring equipment safety and regulatory compliance.

\subsection{Other Critical Challenges}

This subsection discusses other critical cross-cutting challenges arising from the rapid expansion of AI data centers, including cybersecurity, decarbonization, and water consumption.

\subsubsection{Cybersecurity}

AI data centers not only present major physical impacts on the power grid, as discussed above, but also introduce emerging cybersecurity risks. Their highly concentrated electricity demand, dependence on programmable power electronics, integration with cloud-based control architectures, and potential participation in grid-interactive operation collectively expand the cyber-physical attack surface. There are several typical cyberattack pathways. First, attackers could manipulate workload orchestration, such as training schedules, batch inference queues, or compute-resource allocation, to create synchronized load surges or sudden load reductions across one or multiple facilities. Second, compromised energy/cooling management systems, UPS/BESS controllers, or microgrid controllers could trigger inappropriate charging, discharging, generator dispatch, mode transitions, reconnection actions, or load response. Third, falsified telemetry, load forecasts, flexibility reports, or demand-response signals could mislead data center and grid operators, resulting in incorrect reserve scheduling, dispatch, or load-management decisions. Fourth, attacks on data center-utility communication channels could delay, block, or falsify dispatch instructions, ride-through coordination, or emergency operating procedures. 

 As a result, cyber intrusions into AI data center operations may translate into large and fast physical power disturbances, potentially causing frequency excursions, voltage violations, reserve depletion, local network overloads, or cascading reliability risks. For instance, in 2024, a large-scale disconnection event occurred in Virginia's Data Center Alley, where a protection system failure caused 60 out of more than 200 data centers to suddenly disconnect from the grid and transition to on-site generators~\cite{McLaughlin_2025}. This abrupt loss of load introduced a large power imbalance, requiring the utility to rapidly curtail generation to avoid cascading outages. Although this event was not caused by a malicious cyberattack, it illustrates the grid stress that can result from sudden and correlated data center load changes.
Beyond transient disturbances, maliciously orchestrated or malfunction-induced workload patterns could also act as sources of \emph{forced oscillations}, periodically modulating AI data center demand in ways that resonate with the grid's natural modes. Such oscillatory load injections can amplify inter-area oscillations, reduce damping margins, and, in extreme cases, trigger widespread instability~\cite{sun2018cyber}. The controllability and programmability of AI workloads make them particularly susceptible to being exploited as virtual power perturbation sources.

A cyber-physical defense framework tailored to AI data centers is required to mitigate these risks. Key measures include (\romannumeral1) secure and authenticated communication among data center operators, utilities, aggregators, and system operators; (\romannumeral2) network segmentation between enterprise IT, AI workload management, and operational technology systems; (\romannumeral3) real-time anomaly detection that jointly monitors cyber indicators and physical responses, such as workload patterns, power consumption, ramp rates, voltage/frequency dynamics, and oscillatory signatures; (\romannumeral4) validation and redundancy for load forecasts, flexibility offers, and demand-response signals; and (\romannumeral5) operational safeguards, such as ramp-rate limits, staged reconnection, ride-through requirements, and fail-safe UPS/BESS control logic. Coordinated incident-response protocols between data center and grid operators are also needed to prevent local cyber incidents from propagating into broader grid reliability events. 
Cybersecurity at the AI data center-power grid nexus remains an emerging research area. Future work is needed to develop attack-impact quantification methods, cross-domain threat models, validated cyber-physical datasets, and coordinated defense mechanisms involving data center operators, utilities, and system operators.

\subsubsection{Decarbonization} 

According to the IEA report~\cite{iea2024energyai}, carbon emissions from the electricity use of global data centers reached 180 million tons (Mt) in 2024 and are projected to rise to 300 Mt by 2035, raising concerns that the rapid growth of AI facilities could further exacerbate climate change~\cite{Gibney2022Jul}. AI data centers generally do not produce direct carbon emissions during normal operation, except from on-site backup generators such as diesel or natural gas units that are typically activated during outages or emergency operation. Instead, their dominant carbon footprint is indirect and associated with electricity consumption\footnote{Accordingly, the Greenhouse Gas (GHG) Protocol~\cite{world2014scope2} defines Scope 1 and Scope 2 emissions to distinguish between direct and indirect emissions for carbon accounting.}, with carbon emission intensities determined by the generation fuel mix and electric power flows of the supplying power grid~\cite{chen2024carbon}; see~\cite{chen2024towards} for a more detailed introduction.

The decarbonization of AI data centers faces several technical and economic bottlenecks. First, AI data centers require a continuous and highly reliable electricity supply, whereas renewable resources such as wind and solar are variable and location-dependent. This creates a temporal mismatch between real-time data center demand and the availability of carbon-free generation. Achieving hourly or real-time carbon-free energy matching therefore requires firm low-carbon resources, long-duration energy storage, flexible workload scheduling, or geographic load shifting, all of which increase operational and infrastructure complexity. Second, many regions with rapidly growing data center loads face transmission congestion, limited clean-generation interconnection capacity, and long grid-upgrade timelines, which constrain the physical deliverability of renewable or low-carbon electricity to data center sites. Third, firm low-carbon supply options, such as advanced nuclear, geothermal, hydrogen-based generation, or long-duration storage, remain costly or are not yet widely deployable at the scale and speed required by AI data center growth. Fourth, annual renewable power purchase agreements (PPAs)~\cite{kansal2018introduction} and renewable energy certificates (RECs)~\cite{lau2008bottom} can reduce market-based emissions accounting, but they may not ensure that data centers are physically supplied by carbon-free electricity during every operating hour. Hence, there exists a gap between annual clean-energy procurement and real-time grid decarbonization.

At present, data center owners pursue decarbonization primarily through co-located renewable generation, PPAs, RECs, and emerging 24/7 carbon-free energy procurement strategies. For example, Google launched its ``24/7 Carbon-Free Energy'' program~\cite{google247program}, which aims to match data center electricity consumption with carbon-free electricity on a continuous hourly basis. However, realizing such goals at scale remains challenging because it requires not only additional clean generation, but also sufficient transmission capacity, firm low-carbon resources, storage, and operational flexibility to align AI computing demand with low-carbon electricity availability. 
Consequently, the \emph{compute-electricity-carbon
coupling} between data centers and the grid  highlights the importance of jointly considering energy efficiency within data centers, carbon-aware workload scheduling, flexible demand response, clean generation procurement, and the broader decarbonization trajectory of power grids~\cite{chen2024enhance}.

\subsubsection{Water Consumption}

High-density AI computing equipment produces substantial heat, which must be continuously removed to maintain safe operating temperatures and avoid hardware failures.
Many large-scale data centers consume substantial volumes of water daily to dissipate heat, often amounting to tens of thousands or even millions of liters. As indicated in the IEA report~\cite{iea2024energyai}, a 100 MW data center in the U.S. consumes around 2 million liters of water per day on average, equivalent to about 6,500 households. The same report estimates that global water consumption for data centers is currently around 560 billion liters per year and could rise to around 1,200 billion liters per year by 2030. Specifically, 
water consumption in data centers is mainly through evaporative cooling systems and cooling towers, where water evaporation is used to reject heat from chilled-water loops or air-handling systems~\cite{DOE2019CoolingWater,Microsoft2023Cooling,Equinix2024Water}. Additional water use may occur in humidification systems that maintain indoor environmental conditions for IT equipment, as well as in water treatment and blowdown processes that control mineral accumulation in cooling systems~\cite{DOE2019CoolingWater}. Beyond these direct water uses, data centers also contribute to indirect water consumption through the electricity generation used to power computing and cooling loads, particularly in regions where thermal power plants are major generation sources with substantial water consumption for cooling~\cite{Lei2025Jun}.

The water impact of data centers depends not only on total water consumption, but also on where and when water is consumed. Reference~\cite{wu2025not} proposes a framework for evaluating the water impact of computing by accounting for spatial and temporal variations in water stress. In reference~\cite{Lei2025Jun}, the factors influencing data center water use are analyzed, including server efficiency, server utilization, cooling technology, infrastructure efficiency, climate zone, and the water intensity of  electricity supply. The study also finds that there is no single recipe for minimizing water use; instead, optimal outcomes depend on tailored combinations of these factors. 
To reduce stress on water resources, mitigation strategies should be aligned with the specific sources of water consumption. For cooling-related water use, effective approaches include dry heat rejection and pairing direct-to-chip, immersion, or hybrid cooling with low-water heat-rejection systems to reduce reliance on evaporative cooling in water-scarce regions~\cite{Zhang2021CoolingSurvey,Azarifar2024LiquidCooling,Lei2025Jun}. For cooling-tower operation, improved water treatment, cycles-of-concentration management, and the use of reclaimed, recycled, or non-potable water can reduce freshwater withdrawals~\cite{DOE2019CoolingWater}. 
For indirect water use through electricity supply, carbon- and water-aware workload scheduling can shift flexible computing tasks toward times and locations with lower water-stress impacts~\cite{wu2025not,Lei2025Jun}.

 Together with high electricity demand, substantial water consumption is emerging as a critical limiting factor for AI data center expansion, especially during summer peaks when both water availability and power supply are under stress. According to Bloomberg~\cite{Nicoletti2025AIWater}, nearly two-thirds of new U.S. data centers built since 2022 are located in high water-stress regions, such as the Southwest, Texas, and parts of the Midwest, some of which face growing risks of drought and water scarcity. 
 A case study~\cite{Karimi2022WaterEnergyTradeoffs} of two Phoenix-area data centers shows that cooling technologies can create significant water-energy tradeoffs, where lower PUE may be accompanied by higher water usage effectiveness (WUE), highlighting the need to evaluate energy and water performance jointly. Recent work~\cite{11355926} proposes a stochastic-robust computing-electricity-water planning optimization framework that co-optimizes the investment decisions for electricity and water delivery network expansion to support the cost-effective deployment of sustainable data centers. In reference~\cite{11341378}, a joint planning strategy for data centers, distribution networks, and water supply networks is proposed based on a two-stage distributionally robust optimization approach, balancing system economy, low-carbon performance, and robustness.
  As suggested by these studies, integrated frameworks that jointly consider water availability, cooling technology choices, renewable energy integration, regional grid conditions, and flexible workload scheduling are increasingly important when planning and operating future AI data center infrastructure.

\section{Potential Solutions and Opportunities}
\label{sec:solution}

\begin{figure*}[htbp]
    \centering
      \includegraphics[width=0.9\textwidth]{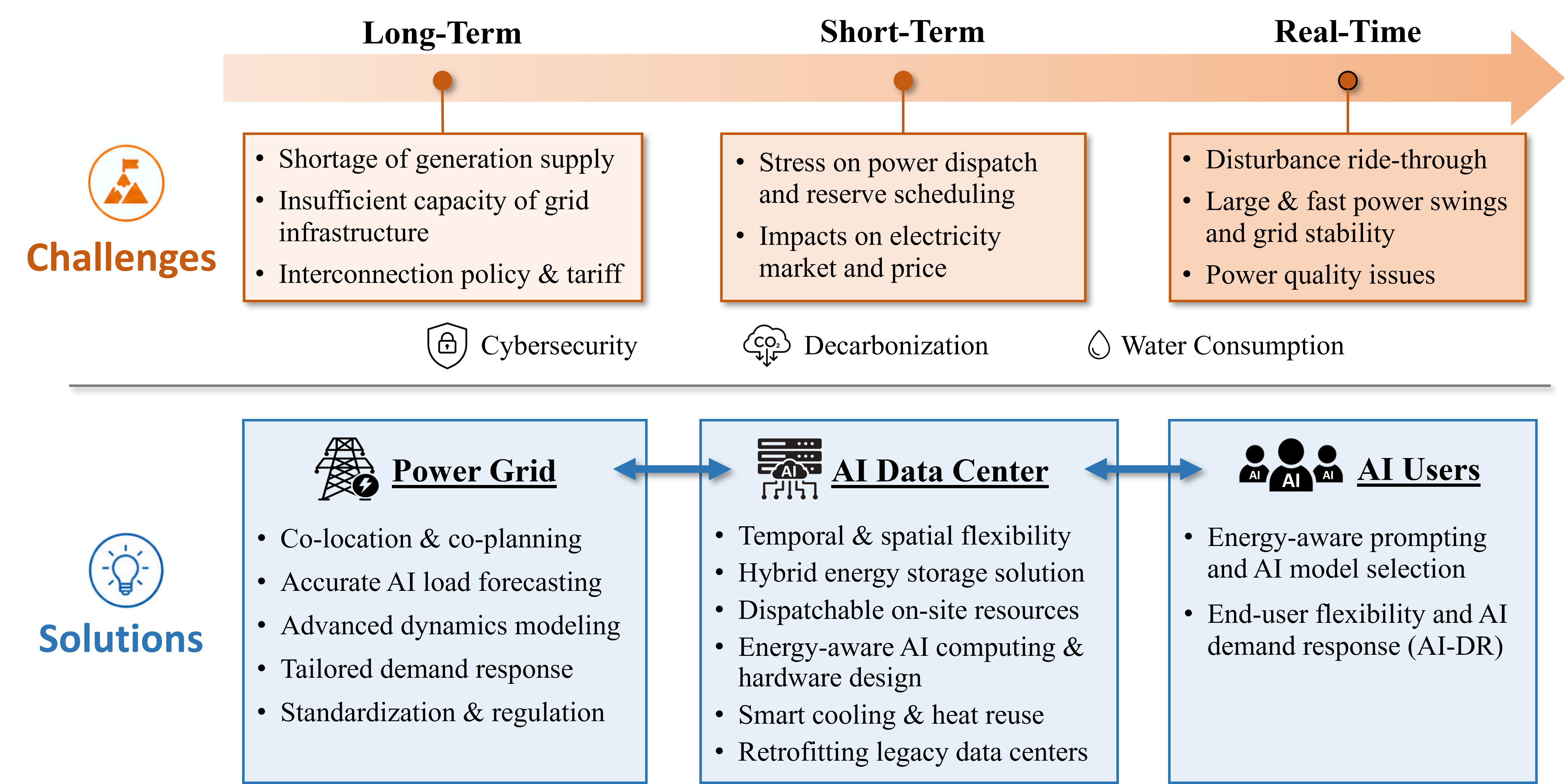}
    \caption{Critical challenges and potential solutions for the large-scale integration of AI data centers into the power grid.}
    \label{fig:impactsolu}
\end{figure*}

In addition to the approaches discussed in Section \ref{sec:gridimpact}, this section presents further solutions and opportunities for addressing the challenges associated with the rapid integration of AI data centers. The discussion is organized from three perspectives: the power grid, AI data centers, and AI end-users. Then,
an overview of existing demonstration projects on computing-electricity coordination is provided, followed by an integrated solution framework and a discussion of practical deployment considerations.
Figure \ref{fig:impactsolu} visualizes the critical challenges discussed in Section \ref{sec:gridimpact} and the potential solutions introduced in this section. 

\subsection{Solutions on Power Grid Side}

\subsubsection{AI Load Forecasting}

Accurate forecasting of AI data center loads, particularly during large-scale model training and inference stages, is critical for grid planning, operation, and control. 
Unlike traditional data center loads, AI workloads exhibit highly irregular temporal patterns, with abrupt transitions between low-utilization and peak-demand periods. 
Effective forecasting requires a comprehensive understanding of AI workload characteristics, including job scheduling policies, training cycle durations, hardware utilization patterns, and the interaction between compute intensity and cooling system demand~\cite{cupelli2018data}. 
A key challenge is that many predictors of AI data center load variations are internal to the data center and are not directly observable by grid operators, such as job queue status, planned training start and stop times, batch inference requests, GPU utilization, and cooling system operating states. Therefore, \emph{collaborative prediction} is useful for improving AI data center load forecasting. Specifically, collaborative prediction refers to a privacy-preserving forecasting framework in which data center operators and grid operators combine complementary information to improve forecasting accuracy and grid situational awareness. Data centers generate internal forecasts using workload telemetry and operational schedules, while grid operators can incorporate external variables such as weather conditions, local load patterns, network congestion, electricity prices, and reserve requirements. The resulting forecasts support planning and operational decisions for both data centers and grid operators. To preserve privacy, for example, data centers can perform forecasting internally and report only aggregated load forecasts, ramp-rate estimates, uncertainty bands, and expected flexibility ranges to grid operators.  A more integrated approach is anonymized feature sharing, in which sensitive workload information is converted into aggregated indicators, such as expected GPU utilization and the probability of peak load events, without revealing proprietary AI workload details.

As a complement, recent advances in machine learning (ML) techniques can be leveraged to enhance AI load forecasting in a more targeted manner~\cite{Ma2025May,mughees2025forecast}. 
These approaches can integrate telemetry data 
from IT infrastructure, job queue statistics, and environmental factors with advanced ML tools to predict short-term fluctuations and identify potential peak demand events before they occur.
Several useful ML techniques include: (\romannumeral1) time-sequence learning methods, such as long short-term memory (LSTM) networks \cite{lindemann2021survey}, temporal convolutional networks (TCNs) \cite{he2019temporal}, transformer-based models \cite{zeng2023transformers}, and structured state-space sequence models (Mamba)~\cite{gu2023mamba}, can capture short-term temporal dependencies and abrupt ramping patterns in AI computing loads; (\romannumeral2) context-aware learning \cite{chattopadhyay2024context} can improve forecasting accuracy by incorporating not only historical power consumption, but also workload-related features such as scheduled training windows, batch inference requests, GPU utilization, server temperature, cooling system states, and weather conditions; (\romannumeral3) federated learning \cite{taik2020electrical} offers a privacy-preserving approach for collaborative forecasting, allowing multiple data centers or data center-grid operator partnerships to train shared forecasting models without directly exchanging proprietary workload data.
Recent studies on AI data center and cloud/HPC load forecasting show that incorporating workload telemetry and hybrid statistical-ML models can improve short-term prediction accuracy relative to conventional autoregressive models~\cite{li2023shortterm,mughees2025forecast}. As a result, rather than being treated as a generic time-series prediction task, AI load forecasting should be formulated as a workload-aware, privacy-preserving, and collaborative forecasting problem.

\subsubsection{Dynamic Modeling of AI Data Centers and Available Open-Source Models}

High-fidelity modeling of the electrical dynamics of AI data center facilities is essential for assessing their grid disturbance ride-through capabilities, local power quality impacts, and interactions with power system stability and reliability. As inherently power-electronics-based loads, AI data centers exhibit dynamic behaviors governed by internal power converter controls, UPS operation, protection logic, cooling system dynamics, and the temporal characteristics of computational workload processing. Because AI data centers are emerging large loads, their dynamic modeling, validation, and testing remain under development and constitute an active research area in both academia and industry. The recent North American Electric Reliability Corporation (NERC) guidance~\cite{NERC2026LargeLoadFAQ} also emphasizes that existing dynamic load models have limitations in accurately representing emerging large loads such as data centers, highlighting the need for improved modeling methodologies. 

To represent the fast and high-frequency transient behaviors of AI data centers, electromagnetic transient (EMT) modeling is required to capture fast converter controls, waveform-level phenomena, unbalanced conditions, detailed protection and mode-transition behavior.  
Meanwhile, phasor-domain models and reduced-order equivalents remain necessary for large-scale, wide-area power system studies, where fully detailed EMT simulation may be computationally impractical. The  Energy Systems Integration Group (ESIG)~\cite{ESIG2026LargeLoadModeling} report provides a comprehensive overview and recommendations for modeling of large loads in power system dynamic studies. 
Several pioneering efforts have recently advanced the EMT dynamic modeling of AI data centers and other large electronic loads. The joint work between Texas A\&M University and ERCOT developed open-source, high-fidelity PSCAD-based EMT models for large AI data center loads \cite{Wang2026AIDCPSCAD} and crypto-mining loads \cite{Wang2025CryptoMinerPSCAD}, involving key power electronic converters, UPS systems, computing loads, cooling loads, control mechanisms, ride-through behavior, and protection and reconnection logic~\cite{ERCOTLargeLoadModeling}. In parallel, the Pacific Northwest National Laboratory  developed a generic EMT model library for data centers that provides component-level and system-level modeling structures for evaluating the dynamic impacts on power grids~\cite{RossFollum2025PNNL}. In terms of phasor-domain modeling, the Power Electronic Reconnection and Ceasing (PERC) model~\cite{Weber2025PERC1} has been introduced to represent the ceasing and reconnection behavior of large electronic loads.
In \cite{zhao2026dynamic}, a
component-informed dynamic model of data-center power-delivery chains in the positive-sequence phasor domain is presented. 
Reference \cite{jimenez2025dynamic} proposes a data center dynamic model that takes into account the cyclic load behavior and disconnection/reconnection logic. 
Moreover, complementary approaches include hierarchical and hybrid modeling frameworks that combine high-fidelity EMT models for critical components or the data center facility and reduced-order equivalents for the broader grid with lower computational burden~\cite{plumier2016co}. 
This hybrid structure balances computational efficiency with modeling fidelity and is commonly recommended for large-scale grid-level dynamic studies~\cite{ESIG2026LargeLoadModeling}.

\subsubsection{Tailored Demand Response Program Design}

To effectively harness the flexibility of AI data centers while minimizing adverse operational impacts, grid operators need to design demand response (DR) programs tailored to the unique characteristics of these loads. Conventional price-based DR mechanisms, while widely used in residential and industrial contexts, may introduce new challenges when applied directly to large-scale AI loads. For instance, inference or training tasks deferred in response to high electricity prices may be rescheduled simultaneously once prices fall, leading to secondary demand peaks or rebound effects that stress the grid. Moreover, AI data center operators can be relatively insensitive to electricity prices. As noted in~\cite{crozier_potential_2025}, these facilities have high sunk capital costs, and the workloads they run are highly lucrative. As a result, the opportunity cost of deferring training or inference tasks can outweigh potential savings from energy arbitrage. Therefore, relying on price signals alone may be insufficient to drive effective load shifting. 

One promising direction involves incentive-based, penalty-based, and contract-based DR mechanisms, in which AI data centers commit to predefined flexibility capabilities in exchange for direct compensation, reduced demand charges, priority grid services, or other benefits~\cite{oconnell2014benefits,cupelli2018data,zhang2020hpc}. Such contracts may specify the maximum curtailable load, response duration, ramping limits, recovery constraints, and acceptable service-quality impacts. They should also distinguish between delay-tolerant workloads, such as model training and batch inference, and latency-sensitive workloads, such as real-time inference. Prior studies have shown that data centers can provide DR through multiple flexibility sources, including delay-tolerant IT workload scheduling, battery storage, and cooling system control, while maintaining operational constraints and quality-of-service requirements~\cite{cupelli2018data,zhang2020hpc,Lukawski2019DemandResponse}. 
Recent industry activities have further confirmed the practical relevance of this approach. 
For example, Google has signed contractual agreements with utilities in Indiana and Tennessee to reduce power consumption associated with AI workloads during periods of grid stress~\cite{reuters2025googleDR}. In addition, coincident peak mechanisms, such as ERCOT’s Four Coincident Peak (4CP) framework, allocate transmission-related costs based on a customer’s contribution to system peak demand during the highest-demand intervals in the summer months, thereby incentivizing large loads such as data centers to reduce or shift consumption away from peak periods~\cite{Lukawski2019DemandResponse,ercot4cp}.

\subsubsection{Standardization and Regulation}

As the scale and concentration of AI data centers increase, specific standards and regulatory frameworks become necessary to manage their grid-interfacing behavior and preserve power system reliability. One key task is the specification of fault ride-through capabilities. Similar to the requirements imposed on renewable energy resources under IEEE Standard 1547 and European grid codes, AI data centers may be required to remain connected for a certain time during short-duration voltage or frequency disturbances. For instance, proposed specifications for data center UPS systems suggest maintaining operation down to 50--70\% of nominal voltage with resynchronization capability within one second~\cite{turn0search3}. In addition to fault ride-through, dynamic performance requirements should address voltage and frequency tolerance ranges, reactive power capability, harmonic emissions, ramp-rate limits, staged reconnection, and acceptable load fluctuation levels. Such performance standards are increasingly being considered for large electrical and hybrid loads under evolving interconnection frameworks~\cite{turn0search12}. Moreover, interconnection standards may require AI data center operators to provide utilities and system operators with accurate dynamic models, anonymized workload profiles, and operational telemetry to enable realistic planning and stability analysis~\cite{gridlab2025}. 
There is also growing momentum for applying grid-code-like requirements to large electronic loads, including mandatory response characteristics during contingencies~\cite{turn0search8}. These regulatory and standardization measures are critical to ensuring that AI data centers operate not only as large electricity consumers but also as reliable and supportive grid participants.

A potential roadmap for developing and implementing these standards and regulations includes several stages: (\romannumeral1) policymakers and reliability organizations should define applicability thresholds for large load facilities, such as AI data centers above a specified MW power capacity or with high power-electronics penetration; (\romannumeral2) interconnection processes should require standardized data submissions, including facility load composition, UPS and backup generation characteristics, protection settings, expected ramp rates, anonymized workload profiles, and validated dynamic models; (\romannumeral3) grid codes should establish minimum technical performance requirements, including voltage and frequency ride-through, reactive power support, harmonic limits, ramp rate constraints, and reconnection coordination; (\romannumeral4) compliance procedures should be developed through model validation, hardware-in-the-loop testing, commissioning tests, and periodic re-verification after major changes in computing hardware, UPS systems, or control settings; (\romannumeral5) operational coordination requirements should be specified, including telemetry sharing, event reporting, emergency communication protocols, and coordination of demand response or load-shedding actions with grid operators. Finally, these standards and regulations should be updated periodically as AI workloads, power electronic interfaces, and grid reliability requirements evolve.

\subsection{Solutions on AI Data Center Side}\label{sec:centersolu}

\subsubsection{Leveraging Temporal and Spatial Flexibility of AI Workloads for Grid-Aware Operation} 

AI workloads with temporal and spatial flexibility can be rescheduled or shifted across locations without compromising service quality, thereby providing valuable flexibility to support grid operation. Specifically, 
\emph{temporal flexibility} refers to the ability to shift AI workloads over time without compromising service quality or user experience. For example, non-urgent computational tasks, such as large-scale model training or batch inference, can be scheduled during off-peak hours or periods of low grid stress and high renewable energy availability. This helps to flatten load curves, enhance grid utilization, lower operating costs, and reduce carbon emissions. \emph{Spatial flexibility} refers to the capability to allocate AI workloads across multiple data centers in
different geographic locations. 
This allows AI workloads to be rerouted to facilities in regions with cleaner electricity, lower prices, or less grid congestion. 
Together, temporal and spatial flexibility offer powerful levers for optimizing the alignment between AI demand and power system conditions, helping to enhance grid reliability, reduce carbon emissions, and lower operational costs. 
 Prior research shows that temporally shifting AI workloads can reduce operational costs by up to 12\% and carbon emissions by approximately 10\%~\cite{frontiers2024temporal,sciencedirect2024spatial}. Real-world implementations \cite{radovanovic2022carbon} are increasingly supported by energy-aware scheduling platforms and demand-sensing technologies. Moreover, studies based on locational marginal emissions highlight that even modest spatial shifting of loads can yield significant emission reductions~\cite{arxiv2020marginalemission}.

\subsubsection{Hybrid Energy Storage Solutions}
Energy storage systems are increasingly essential for mitigating the fast and large power fluctuations of AI data centers.
However, individual energy storage technologies often fall short of meeting the full spectrum of performance requirements, including rapid response, adequate power and energy capacity, long cycle life, and cost-effectiveness. As a result, hybrid energy storage systems (HESS), which integrate multiple storage technologies with complementary operational characteristics, have emerged as a promising solution to address these diverse demands.
Specifically, \emph{supercapacitors} \cite{zheng2016hybrid} offer sub-second response times and exceptionally high cycle life, making them well-suited for absorbing ultra-short-term power spikes at the server or rack level. 
\emph{Batteries}, particularly lithium-ion and emerging alternatives such as sodium-ion, provide higher energy density and longer-duration buffering, which make them suitable for load shifting, peak shaving, and backup at the facility level \cite{guo2021integrated}. \emph{Flywheels} \cite{amiryar2017review} deliver medium-duration, high-power buffering with high round-trip efficiency, making them effective for smoothing aggregated power fluctuations and providing short-term ride-through support at the cluster level. A potential hybrid configuration, in which supercapacitors manage millisecond-to-second transients, flywheels address second-to-minute ramps, and batteries handle minute-to-hour variations, offers a balance of technical performance, lifecycle efficiency, and overall cost-effectiveness. As AI workloads continue to grow in intensity and variability, the deployment of hybrid storage architectures, coupled with smart energy storage management systems, will play a critical role in enabling sustainable, reliable, and grid-friendly AI data center operations.

 \subsubsection{Dispatchable On-Site Energy Resources}

AI data centers can leverage on-site energy resources to support grid operations and enhance power system flexibility. As introduced in Section~\ref{sec:powerinf}, backup generators, on-site renewables, UPS systems, and other energy storage systems are primarily deployed to ensure reliable operation during grid disturbances. Nevertheless, when coordinated with grid operators, these flexible resources can also provide dispatchable demand-side support. For example, backup generators can temporarily reduce net grid demand during peak periods or emergency conditions, while UPS and battery systems can provide fast-response services~\cite{aljbour2024powering}, including peak shaving, ramp-rate mitigation, and frequency regulation. However,
the use of backup generators for grid-support services must be carefully managed to avoid compromising data center reliability, increasing emissions, or violating operational constraints. These resources are therefore most effective when integrated into smart energy management systems that coordinate backup generation, on-site renewables, energy storage, and AI workload scheduling. Such coordination can help balance operational reliability with grid-support objectives, positioning AI data centers not only as large electricity consumers but also as active grid participants that enhance system reliability.

\subsubsection{Energy-Aware AI Computing and Hardware Design} AI computing accounts for the largest share of electricity consumption in AI data centers, making advances in computing efficiency and management crucial for reducing their energy footprint and stabilizing power demand. Several complementary strategies are outlined below:

(\romannumeral1) \emph{Energy-aware AI computing}: Techniques such as energy-aware training (EAT)~\cite{lazzaro2023minimizing} enable models to activate only the neurons and connections necessary for a given task, effectively disabling unused components and lowering power use. Additional savings can be achieved through dynamic voltage and frequency scaling of CPUs and GPUs. For example, reference \cite{crozier_potential_2025} reports that under-clocking a single NVIDIA A100 GPU for a specific application reduced power consumption by 40\% while decreasing performance by only 22\%. This demonstrates that certain jobs could be executed at lower frequencies over a longer duration to reduce overall energy use. Moreover, advanced power smoothing and management techniques at the software and GPU levels can be developed and deployed to mitigate power swings induced by AI workloads \cite{choukse2025power}, thus reducing risks to grid stability.

(\romannumeral2) \emph{Development of lighter, task-specific AI models}: Smaller, task-specific AI models can match the accuracy of large foundation models while consuming substantially less power. Techniques such as pruning (removing redundant parameters), quantization (reducing numerical precision), and knowledge distillation (transferring knowledge from a large model to a smaller one) lower computational demands without compromising output quality. For example, TinyBERT achieves over 90\% lower energy consumption compared to full-size BERT models \cite{paula2025comparative}. {\color{black}
These approaches also facilitate the deployment of edge AI computing on edge or mobile devices while yielding significant energy savings. Specifically, as AI foundation models become more mature and widely deployed, inference workloads may account for a major share of total AI computing demand. In this context, edge computing can serve as a complementary paradigm by offloading selected inference workloads from hyperscale data centers, reducing latency, preserving privacy, and partially alleviating centralized grid impacts. It is well-suited for latency-sensitive, localized, and geographically distributed applications, such as autonomous vehicles, Internet of Things (IoT) devices, distributed energy resources, industrial monitoring, and smart-city systems. By processing selected lightweight AI tasks locally, edge AI computing can complement hyperscale AI data centers, reduce data transmission requirements, distribute electricity demand across more locations, and moderate the concentration of load growth in major data center hubs.
}


(\romannumeral3) \emph{Energy-efficient hardware design}: The deployment of energy-efficient hardware is necessary for reducing the carbon and energy footprint of AI data centers. Advances in processor architecture can substantially improve performance-per-watt, enabling higher computational throughput at lower power costs. Domain-specific accelerators, low-power CPUs/GPUs, and AI-optimized hardware such as Google’s TPUs are central to this effort. For instance, Google’s TPU v4-based supercomputer has demonstrated notable speed and efficiency gains over NVIDIA A100 GPUs~\cite{google2023tpuv4}. Moreover, AI itself can be leveraged to optimize the performance and energy efficiency of its supporting infrastructure, as illustrated in~\cite{bolon2024green}. Co-optimizing AI model architectures with underlying hardware, combining techniques such as pruning, quantization, and hardware-aware model design, offers further potential for achieving substantial energy savings without sacrificing accuracy. 

\subsubsection{Smart Cooling and Heat Reuse}
Cooling facilities account for the second-largest share of electricity consumption in AI data centers. 
Recent research and industry trends emphasize several advanced strategies: (\romannumeral1) \emph{Hybrid air-liquid cooling systems}: combining air-based airflow management (e.g., hot/cold aisle containment, in-row cooling, raised-floor systems) with liquid approaches (e.g., direct-to-chip) to optimize energy, complexity, and adaptability. These systems help manage thermal loads from dense AI clusters while reducing energy usage compared to air-only or liquid-only designs \cite{TakingHybrid2024, GuideToCooling2025}. (\romannumeral2) \emph{Immersion and direct liquid cooling}: immersing servers in dielectric fluids or using pumped liquid-cooling loops offers dramatically improved heat transfer, higher rack density, and up to 50\% energy savings compared to air cooling, though challenges remain around maintenance and cost \cite{ ImmersionCooling2025,mohammedPerformanceImprovementHighdensity2025}.  (\romannumeral3) \emph{AI-driven adaptive and predictive cooling control}: leveraging reinforcement learning (RL) \cite{chen2022reinforcement}, deep neural networks, or other ML models to anticipate thermal load changes and dynamically tune cooling systems for maximum efficiency. These approaches have achieved 14-21\% energy savings in live data center deployments without violating safety constraints \cite{Zhan2025DC21}. Google’s commercial cooling experiments using RL yielded 9-13\% energy reductions \cite{Luo2022RL}. (\romannumeral4) \emph{Renewable-powered and free (economizer) cooling}: leveraging ambient air (free cooling) where climates permit, and powering cooling infrastructure with on-site renewables (solar, wind) to reduce carbon intensity. These approaches support sustainable operations while attenuating grid demand.

Since most of the electricity consumed by AI computing is ultimately converted into heat, another promising direction is \emph{heat reuse}, in which waste heat from AI servers is recovered and supplied to nearby thermal loads, such as district heating networks, buildings, greenhouses, or industrial processes. This approach is especially attractive for high-density AI data centers using liquid cooling, because higher-temperature coolant streams provide more useful recoverable heat than conventional air-cooling systems~\cite{Zimmermann2012Aquasar,Meyer2013IDataCool}. Heat reuse can improve system-level energy utilization and reduce the need for separate heating energy, but its feasibility depends on coolant temperature, heat-exchanger design, distance to heat users, seasonal heating demand, local thermal-network infrastructure, and economic incentives. To this end, future AI data centers can be designed as heat-recovery-ready facilities by co-optimizing cooling loops, thermal storage, and local heat-demand integration.

\subsubsection{Retrofitting Existing Data Centers for AI Workloads}

To support the rapid growth of AI workloads, many data center owners are choosing to retrofit and upgrade existing facilities to accommodate high-density AI computing. Compared with developing entirely new facilities, retrofitting can shorten deployment timelines, reduce upfront development uncertainty, and leverage existing land rights, fiber connectivity, security infrastructure, and grid interconnections. As a consequence, upgrading legacy data centers has emerged as a practical near-term pathway for expanding AI capacity amid growing constraints in grid availability, equipment lead times, land acquisition, and permitting. 

However, retrofitting existing data centers for AI workloads is technically demanding, as AI servers introduce substantially higher power density, stronger thermal gradients, and more dynamic power profiles than conventional enterprise or cloud workloads \cite{avelar2023ai}. Therefore, retrofit projects require a systematic assessment of structural, electrical, mechanical, thermal, and operational constraints \cite{CadenceRetrofittingOlderDataCentersAI2026,avelar2023ai}. 
From a structural perspective, facilities must be evaluated for floor loading, rack weight, battery weight, ceiling support, overhead busway installation, liquid-cooling manifolds, and congestion in raised-floor or overhead distribution spaces. From an electrical perspective, upgrades are required for power transformers, medium-voltage switchgear, UPS modules, power distribution units, backup generators, protection coordination, short-circuit ratings, grounding, harmonic mitigation, and power-quality compliance. From a thermal perspective, conventional air-cooling systems may be insufficient for high-density AI racks, requiring direct-to-chip liquid cooling, rear-door heat exchangers, coolant distribution units, upgraded chilled-water loops, additional pumping capacity, leak detection, and enhanced heat-rejection systems \cite{SchneiderBrownfieldDataCenterModernizationAI2026,wang2026revolutionizing}. These upgrades often need to be implemented in live operating environments, where construction sequencing, limited outage windows, redundancy preservation, commissioning procedures, and risk mitigation are critical. Consequently, retrofit-aware planning should jointly consider power delivery, cooling capacity, reliability requirements, staged construction, and workload deployment, making it an important research and engineering problem for near-term AI data center expansion.

\subsection{Solutions on AI End-User Side}

\subsubsection{Energy-Aware Prompting and AI Model Selection}

AI application developers and 
end-users can contribute to reducing the AI electricity footprint through energy-aware application development and informed usage decisions. 
For end-users, one important approach is the adoption of energy-efficient prompting strategies and the selection of appropriately scaled models. 
Recent studies have shown that longer, more verbose prompts, while potentially improving model interpretability or output alignment, 
can significantly increase the number of tokens to be processed and thus substantially raise the computational workload and energy consumed per query~\cite{rubei2025prompt,adamska2025green}. 
Moreover, fine-tuned, task-specific lightweight AI models often yield comparable performance to general-purpose foundation models but at a small fraction of the computational and energy cost~\cite{oconnell2014benefits}. 
{\color{black}To implement these approaches, AI service providers can develop system-level prompt optimization and model-routing tools that automatically refine user prompts, remove redundant context, reduce unnecessary token usage, and select task-appropriate models while preserving service quality. In this way, end-users can achieve the same task objectives with lower computational and energy costs, without needing to manually evaluate the energy implications of each query. Moreover,
the practical adoption of energy-aware prompting and model selection also requires appropriate incentive structures. 
AI service providers can facilitate energy-aware behavior through transparent energy or carbon metrics, energy-efficiency labels, estimated energy use per query, pricing incentives, usage credits, delayed-processing discounts for non-urgent tasks, and opt-in low-carbon service modes. Recent initiatives, such as the AI Energy Score project~\cite{turn0search14}, aim to improve transparency by benchmarking the energy cost per query across different AI models. Making such metrics available to end-users can support more informed usage decisions and incentivize developers and service providers to adopt energy-aware design considerations.
}



\subsubsection{User Flexibility and AI Demand Response}
In practice, many end-users of AI applications exhibit a degree of flexibility in their usage behaviors, as they often do not need immediate AI responses to their queries. For example, a user submitting a complex AI task late at night may 
not need the results until the following morning. This inherent temporal flexibility in the AI demand of end-users offers a valuable opportunity to better align query processing with power grid operational needs. 
Building on this insight, a novel concept, termed ``\emph{AI Demand Response}" (AI-DR), is proposed in this paper for harnessing user-side AI workload flexibility, analogous to electric demand response \cite{oconnell2014benefits,chen2021online} in power systems.

Similar to electric demand response programs, 
potential AI-DR mechanisms can be broadly categorized into \emph{incentive-based} and \emph{price-based} programs.
 In incentive-based AI-DR programs, end-users 
 are incentivized through lower subscription fees, discounted inference rates, usage credits, or
access to premium model tiers in exchange for accepting minor service compromises \cite{crozier_potential_2025}. Such flexibility may include bounded response delays, use of lower-power or lower-tier hardware, energy-aware model selection, or query processing during periods of low grid stress, low electricity prices, or high renewable generation. 
In contrast, 
price-based AI-DR programs employ time-varying pricing schemes, such as real-time pricing, time-of-use pricing, or critical peak pricing \cite{oconnell2014benefits}, to guide users to shift non-urgent AI usage away from peak periods. By exposing end-users to dynamic AI usage pricing signals, service providers (AI data centers) can shape AI inference workload patterns in a grid-aware and energy-efficient manner. 
When aggregated across a large and diverse user base, flexible AI demand can provide substantial controllable load capacity, enabling service providers to support grid reliability, reduce operating costs, improve energy efficiency, and lower carbon emissions while preserving the quality of service.

The practical implementation of AI-DR requires careful market and mechanism design, with explicit consideration of both service-level constraints and hardware-health impacts. Key factors include incentive compatibility, participation rules, service-quality guarantees, privacy protection, user acceptance, hardware reliability, and coordination with grid operators. In particular, AI-DR programs should not be applied uniformly to all AI workloads. For example, enterprise AI tenants often operate under strict Service Level Agreements (SLAs), which specify requirements on response latency, throughput, availability, and tail-latency performance. 
Existing studies on AI inference serving have shown that meeting request-level latency and service-level objectives is a central design constraint for production AI systems \cite{GujaratiClockwork2020, RomeroINFaaS2021}. 
Therefore, service-aware AI-DR mechanisms should distinguish between latency-sensitive requests that require immediate service and delay-tolerant workloads that can be shifted in time, processed with lower-power hardware, assigned to smaller task-appropriate models, or routed to lower-carbon computing resources. Service providers may also need to define response-delay limits, service-quality guarantees, compensation levels, opt-in participation options, and transparent energy or carbon metrics so that users can understand the tradeoffs associated with different service choices.

 \subsection{Demonstration Projects on  Computing-Electricity Nexus}

A growing number of international demonstration projects and industry initiatives have been dedicated to computing-electricity coordination.
 These efforts can be broadly grouped into three pathways: (\romannumeral1) \emph{Workload flexibility}. Google has demonstrated carbon-aware and demand-response computing by shifting delay-tolerant workloads across time and locations~\cite{radovanovic2022carbon}. These efforts include earlier collaborations with OPPD, Elia, Centrica Energy, and Taiwan Power Company, as well as recent utility agreements that provide approximately 1 GW of data-center demand-response capacity during grid stress events~\cite{googleBelgiumFlexibleDR}. In addition, recent field demonstrations in Phoenix, Arizona, have shown that AI cluster workload orchestration can reduce power consumption during peak grid events while maintaining AI quality-of-service  guarantees~\cite{colangelo2026ai}. (\romannumeral2) \emph{Grid-interactive power infrastructure}. Microsoft and Eaton have demonstrated the use of data-center UPS batteries to provide fast frequency response and other grid services in Ireland and the Nordic region, showing that backup power assets can be converted into active flexibility resources without compromising data center reliability~\cite{paananen2021grid}. 
 The Electric Power Research Institute (EPRI)'s DCFlex initiative extends this direction by coordinating data-center operators, utilities, and technology providers in field demonstrations across the U.S. and Europe to test flexible load control, interconnection support, and backup-power-to-grid-service strategies~\cite{epri2025dcflex,epri2026dcflex}. (\romannumeral3) \emph{Coordinated infrastructure planning}. In China, the national ``Eastern Data, Western Computing'' initiative coordinates data center deployment with regional energy-resource availability by shifting suitable computing capacity from eastern demand centers to western regions with more abundant land and energy resources. The recently released \emph{Action Plan for Promoting the Mutual Empowerment of Artificial Intelligence and Energy}~\cite{ndrc2026aienergy} presents China’s national strategy for integrating AI development with energy system modernization. In addition, as the world's first commercial offshore wind-powered subsea data center,
 the Shanghai Lingang Underwater Data Center demonstrates an integrated computing-electricity-cooling pathway by coupling data center operation with offshore wind power and seawater cooling~\cite{wired2025lingangUDC}. Together, these projects suggest that computing-electricity coordination is moving from conceptual design toward practical implementation.

\subsection{Integrated Solution Framework and Practical Deployment}

Although the potential solutions discussed above are presented from the separate perspectives of the power grid, AI data centers, and AI end-users, these stakeholders are closely coupled through a hierarchical demand chain. 
At the AI end-user layer, prompt design, model selection, and service-quality preferences affect the timing, volume, and computational intensity of AI requests. At the AI data center layer, these requests are translated into computing, cooling, and power infrastructure demands through workload scheduling, hardware utilization, thermal management, and on-site energy resource operation~\cite{cupelli2018data}. At the grid layer, the resulting electricity demand interacts with generation dispatch, reserve scheduling, transmission congestion, electricity prices, and time-varying carbon intensity~\cite{chen2024enhance,oconnell2014benefits}. 
Therefore, these three layers are inherently interdependent. User-side flexibility, such as delayed processing or energy-aware model selection, can expand the scheduling space available to data centers. Data-center-side workload and energy management can translate computational flexibility into grid-responsive load behavior through workload shifting, cooling control, and energy storage operation~\cite{cupelli2018data,zhang2020hpc,sciencedirect2024spatial}. In turn, grid-side price signals, carbon signals, reliability alerts, and incentive mechanisms can guide both data center operation and user-facing service design~\cite{radovanovic2022carbon,google247program}. Hence, it is critical to unify the potential solutions across these layers within an integrated multi-layer framework to coordinate their interactions and maximize their effectiveness. Such a framework should include three key functions: (\romannumeral1) information exchange, including load forecasts, flexibility availability, grid conditions, and carbon intensity; (\romannumeral2) incentive alignment, including differentiated pricing, demand response contracts, and user-facing energy-aware service options; and (\romannumeral3) coordinated control, such as 
energy-aware workload shifting and grid-aware hybrid energy storage operation. Under this framework, AI end-users provide flexible demand preferences, AI data centers act as flexibility aggregators and controllable large loads, and grid operators provide operational signals and reliability constraints. Such cross-layer coordination can help reduce peak demand, enhance grid reliability, lower carbon emissions, and preserve AI service quality.

Moreover, the practical deployment of the solutions discussed above depends critically on their economic feasibility and cost-benefit performance. This challenge is  significant for joint data center-grid planning, since both AI data center deployment and grid infrastructure expansion require substantial capital investment, yet they often follow different planning horizons, financing structures, and regulatory processes. AI data centers can often be planned and constructed within a few years to meet rapidly growing computing demand, whereas transmission upgrades, substation expansion, and new generation interconnection may require substantially longer timelines for planning, permitting, cost recovery approval, financing, and construction. This mismatch creates a coordination gap: data center developers require timely power availability, while utilities and system operators must ensure that network investments are prudent, reliable, and economically justified. 
In addition, future AI electricity demand remains highly uncertain due to rapid changes in model architectures, hardware efficiency, inference demand, workload management strategies, and business decisions by AI service providers. This uncertainty makes it difficult to determine when, where, and at what scale grid upgrades should be built. 

Cost allocation and risk sharing are also major barriers, as utilities, existing ratepayers, data center developers, generation owners, and regulators may disagree on who should bear the costs of transmission and distribution upgrades, dedicated generation, interconnection facilities, and reliability services. If projected AI demand does not materialize, some infrastructure investments may become underutilized or stranded; conversely, if investment is delayed, grid constraints may slow AI data center deployment and increase reliability risks. These challenges highlight the need for adaptive planning frameworks, transparent interconnection studies, phased investment strategies, and regulatory mechanisms that align incentives and allocate costs and risks fairly among stakeholders.


Furthermore, to facilitate effective management and standardization, it is important to develop robust, validated, and comprehensive metrics for AI data centers to accurately reflect their grid-friendliness and sustainability, beyond traditional indicators such as PUE and WUE. Such metrics should better capture the grid-interfacing behavior and sustainability performance of AI data centers, including their flexibility, grid reliability impacts, carbon footprint, water use, and interactions with local grid conditions. Developing such metrics remains an important direction for future research and standardization.

\section{Conclusion} 
\label{sec:conclusion}

This paper focuses on the growing intersection between AI and energy systems, specifically the electricity demand of AI data centers and their impacts on the power grid. 
It highlights that the electricity use of AI data centers is largely driven by computing and cooling loads, with distinct characteristics and patterns across model preparation, training, fine-tuning, and inference stages.
From the perspective of the power grid, the large-scale integration of AI data center loads introduces multi-faceted challenges across multiple timescales, ranging from long-term planning and interconnection, short-term operation and markets, to
real-time dynamics and stability. Addressing these challenges requires collaborative, integrated solutions involving grid operators, data centers, and end-users, such as advances in AI load forecasting and dynamic modeling, standardization and regulation, grid-aware scheduling, energy-efficient hardware, and energy-conscious AI usage.

Looking ahead, the co-development of power grids and AI data centers will be essential for supporting future energy system infrastructure while sustaining the rapid pace of AI innovation in a reliable and sustainable manner. In particular, several key directions are emphasized:
\begin{itemize}
    \item As large AI models become more mature and widely deployed, \emph{inference} workloads associated with broad downstream applications are expected to account for an increasing, and potentially dominant, share of AI data center electricity demand in the future. Compared with large-scale AI training facilities, inference-oriented data centers may be smaller in individual capacity, more geographically distributed, more heterogeneous in workload composition, and more closely tied to stochastic user request patterns. As a result, they can exhibit more fluctuating and bursty load profiles, although aggregation across many users and services may smooth some variability at larger spatial or temporal scales.
 Further studies are therefore needed to better understand AI inference load patterns, temporal variability, and dynamic behaviors, thereby informing the design of more effective grid-aware scheduling and coordination strategies.

\vspace{5pt}
\item From a technical perspective, 
harnessing \emph{flexibility} is central to enabling grid-friendly operation of AI data centers, such as enhanced disturbance ride-through capability and mitigation of fast power fluctuations. Modern AI data centers contain multiple sources of flexibility, including energy storage, dispatchable on-site resources, power electronic converter control, cooling system thermal flexibility, flexible AI computing workloads, and flexible AI end-user demand. Intelligent and automated management systems are needed to optimally orchestrate these flexible resources across computing, cooling, power infrastructure, and end-user service layers. Effectively harnessing this flexibility, together with proper standards, controls, and grid-coordination mechanisms, can help transform AI data centers from potentially disruptive large loads into grid-friendly and grid-supportive assets.

\vspace{5pt}
 \item From a market-design and regulatory perspective, future electricity market rules and regulatory frameworks need to address two closely related issues: \emph{fair cost allocation} and \emph{effective incentive design}. First, market rules and tariff structures need to reflect cost-causation principles so that data centers contributing to peak demand, increasing electricity prices, transmission congestion, capacity procurement, reserve requirements, or grid infrastructure upgrades bear an appropriate share of the associated costs. This can help avoid shifting AI-driven infrastructure and reliability costs disproportionately to existing ratepayers. Second, incentive mechanisms can reward AI data centers for providing verifiable workload and energy flexibility, such as shifting delay-tolerant computing tasks, reducing demand during system stress, dispatching on-site energy resources, or using storage to support peak shaving and ramping mitigation. By combining fair cost allocation with well-designed flexibility incentives, electricity markets can better accommodate large AI data center loads while encouraging grid-supportive operation.

\vspace{5pt}
 \item  Planning and operating large-scale AI data centers is not a single-domain problem, but a multidisciplinary challenge spanning computing architectures, AI algorithms, power electronics, grid infrastructure, equipment manufacturing, electricity markets, environmental constraints, policy, economics, and engineering practice. Hence, future frameworks should move beyond isolated optimization and develop integrated co-planning and cooperation approaches for the multi-domain ``\emph{compute-electricity-water-carbon-more}" nexus, where computing demand, grid conditions, water availability, cooling requirements, carbon emissions, market design, policy constraints, economic feasibility, and more are considered jointly. Meanwhile, effective solutions require an open and collaborative community connecting AI developers, data center operators, utilities, grid operators, equipment manufacturers, policymakers, and academic researchers. Such cross-sector collaboration is essential for reducing information gaps, sharing representative data and models, developing common standards and metrics, and accelerating practical pathways toward sustainable, reliable, and grid-friendly AI data center growth.

\end{itemize}

\appendices

\section{Global AI Data Center Loads} \label{app:global}

According to the IEA report \cite{iea2024energyai}, global data centers consumed around 415 TWh of electricity in 2024, accounting for about 1.5\% of total global electricity consumption. The United States was the largest contributor, accounting for 45\% of this demand, followed by China at 25\% and Europe at 15\%. 
Given their dominant shares of global data center electricity consumption, the cases in the United States, China, and Europe are discussed in greater detail below; see \cite{iea2024energyai,kamiya2025data,ember2025grids,icis2024hungerpwr} for more data center load information in other countries and regions.

\subsection{AI Data Center Load in the United States}
The United States is among the largest markets for hyperscale data centers and accounts for a significant share of global AI-related electricity demand. 
U.S. data center energy consumption, which reached 176 TWh (4.4\% of total U.S. electricity) in 2023, is projected to increase to a range of 325-580 TWh by 2028, contingent upon various growth scenarios \cite{shehabi20242024}. McKinsey \& Company projects that U.S. data center load will rise from 25 GW in 2024 to over 80 GW by 2030, largely driven by AI workloads \cite{mckinsey2024datacenters}. Moreover, 
AI data center deployment in the U.S. is highly concentrated, as evidenced by the fact that roughly 80\% of the power demand is located in only 15 states \cite{aljbour2024powering}, with Virginia, Texas, Georgia, and Arizona being leading hubs due to their favorable tax incentives, existing fiber networks, and relatively lower energy costs \cite{enconnex2024power}. Northern Virginia’s ``Data Center Alley" alone accounts for more than 2.5 GW of active capacity, with multiple AI-focused data center campuses under construction. New projects are increasingly targeting Midwestern and Mountain West states such as Iowa, Ohio, and Wyoming, for proximity to renewable energy resources and to diversify geographic risk.

\subsection{AI Data Center Load in China}

China is rapidly expanding its AI data center infrastructure, leading to a commensurate increase in power demand. In 2024, the electricity consumption of data centers in China is estimated to be in the 100-150 TWh range, representing about 1-2\% of the nation's total electricity use, which is projected to rise to between 400 and 600 TWh annually by 2030~\cite{carbonbrief2024china}. To manage this growth, China has embedded data center development within its national digital infrastructure and energy planning strategies. The most prominent example is the ``\emph{Eastern Data, Western Computing}" initiative, which aims to shift suitable computing workloads from demand-intensive eastern regions to western and northern regions with more abundant land, renewable-energy resources, and favorable cooling conditions. Under this initiative, the construction of eight national computing hubs and ten national data center clusters has been approved, representing a state-coordinated effort to align computing-capacity deployment with regional energy-resource endowments and grid development \cite{ndrc2022edwc}. Recent research \cite{zhang2024edwc} suggests that this initiative is expected to reduce emissions from the data center sector by 16-20\% by 2030 via improving access to clean electricity, natural cooling, and lower-cost resources.   
In parallel, China has introduced green data center policies that emphasize facility-level energy efficiency, renewable-electricity utilization, and coordinated infrastructure planning. The 2024 \emph{Special Action Plan for Green and Low-Carbon Development of Data Centers}, jointly issued by China's National Development and Reform Commission, Ministry of Industry and Information Technology, and other central agencies, mandates that by the end of 2025 the national average PUE of data centers should be reduced to below 1.5, while renewable energy utilization should increase by 10\% annually~\cite{MIIT2024GreenDataCenters}. The recently released \emph{Action Plan for Promoting the Mutual Empowerment of Artificial Intelligence and Energy}~\cite{ndrc2026aienergy} presents China’s national strategy for integrating AI development with energy system modernization. This policy emphasizes coordinated planning of AI computing infrastructure and energy resources, aiming to improve green electricity utilization in data centers, strengthen ``compute-power" synergy, and accelerate AI-enabled applications in smart grids, energy optimization, and low-carbon transition pathways.

\subsection{AI Data Center Load in Europe} 


Europe is also experiencing a significant expansion of its AI data center capacity. The total IT load demand for data centers in Europe is projected to increase from approximately 10 GW in 2024 to 35 GW by 2030, with corresponding annual electricity consumption nearly tripling from 62 TWh to over 150 TWh by 2030, which would represent about 5\% of the continent's total electricity use~\cite{reuters2024europe}. Accommodating this growth will require substantial investment in both data center infrastructure and grid electricity supply. 
Several European AI and digital infrastructure initiatives have been proposed to support the data center expansion. For example, the European Commission has launched the InvestAI initiative, which aims to mobilize EUR 200 billion for AI investment, including a EUR 20 billion fund for AI gigafactories~\cite{EuropeanCommission2025InvestAI}. In parallel, the European HPC Joint Undertaking is expanding Europe’s public AI and high-performance computing infrastructure through AI factories and exascale systems, such as the JUPITER supercomputer in Germany, which supports scientific computing and AI model training~\cite{eurohpc2025jupiter}.

European data center growth is closely linked to sustainability and regulatory objectives. The Climate Neutral Data Centre Pact commits participating operators to climate neutrality by 2030 and addresses energy efficiency, carbon-free energy procurement, water conservation, circularity of server equipment, and heat reuse~\cite{climatepact2024}. Northern and Nordic regions are particularly attractive for AI computing facilities because of access to low-carbon electricity, cooler climates, and potential waste-heat utilization. Projects such as OpenAI’s proposed Stargate facility in Norway, planned to run on hydropower with an initial 230 MW capacity expandable to 520 MW, exemplify the interest in siting large AI computing facilities near low-carbon energy resources~\cite{itpro2024stargate}.
Meanwhile, Europe faces significant grid-integration challenges. Data center development is concentrated in established hubs such as Ireland, the Netherlands, Germany, France, and the Nordic countries, where grid capacity, land availability, permitting constraints, and environmental impacts may limit further growth. Some regulators are considering more active management of large data center interconnections; for example, Belgium has explored electricity allocation limits and flexible connection arrangements to prevent speculative projects from monopolizing grid capacity and to manage rising AI-driven demand~\cite{reuters2025belgium}. These developments suggest that Europe’s AI data center expansion will require coordinated planning of computing infrastructure, grid capacity, low-carbon generation, flexible interconnection, and sustainability requirements.

\section*{Disclaimer}

The views expressed in this paper are the opinions of the authors and do not reflect the views of PJM Interconnection, L.L.C. or its Board of Managers, of which Le Xie is a member.

\bibliographystyle{IEEEtran}
\bibliography{IEEEabrv,mybibfile}

\end{document}